\documentclass[11pt]{article}

\usepackage[margin=1in]{geometry}
\usepackage{setspace}
\usepackage{abstract}
\usepackage{longtable}
\usepackage[most]{tcolorbox} \definecolor{navyblue}{RGB}{0,51,102}
\usepackage{pdfpages}

\usepackage[table]{xcolor}
\usepackage{array}
\newcolumntype{C}[1]{>{\centering\arraybackslash}m{#1}}
\newcolumntype{L}[1]{>{\raggedright\arraybackslash}m{#1}}

\usepackage{float}

\usepackage[T1]{fontenc}
\usepackage{newtxtext,newtxmath}
\usepackage{microtype}

\usepackage{tabularray}
\UseTblrLibrary{booktabs}

\usepackage{xcolor}
\definecolor{RoyalNavy}{HTML}{123A73}

\usepackage{titletoc}
\usepackage{tikz}
\newcommand{\customsepnavy}{%
  \par\medskip
  \centerline{\begin{tikzpicture}[baseline=-0.5ex]
    \fill[RoyalNavy!60] (-7.5pt,0) circle (1.2pt);
    \draw[RoyalNavy!60, line width=0.75pt, samples=351]
      plot[domain=0:122.5] ({\x*1pt},{1.35*sin(\x/7*360)*1pt});
    \fill[RoyalNavy!60] (130pt,0) circle (1.2pt);
  \end{tikzpicture}}%
  \medskip\par}
\titlecontents{section}[0pt]
  {\addvspace{0.75em}\filright}
  {\color{RoyalNavy}\bfseries\scshape\large\thecontentslabel.\quad}
  {\color{RoyalNavy}\bfseries\scshape\large}
  {\hspace{0.6em}\titlerule*[0.7pc]{\textcolor{RoyalNavy!35}{\textperiodcentered}}\hspace{0.6em}{\color{RoyalNavy}\bfseries\large\contentspage}}

\titlecontents{subsection}[2.2em]
  {\addvspace{0.22em}\filright\small}
  {\color{RoyalNavy}\thecontentslabel\quad}
  {\color{RoyalNavy}}
  {\hspace{0.6em}\titlerule*[0.9pc]{\textcolor{RoyalNavy!35}{\textperiodcentered}}\hspace{0.6em}{\color{RoyalNavy}\contentspage}}

\usepackage{etoolbox}
\makeatletter
\newcounter{affN}
\newcommand{\affblock}{}
\newcommand{\definst}[2]{\csgdef{aff@txt@#1}{#2}}
\newcommand{\aff@get}[1]{%
  \ifcsname aff@num@#1\endcsname\else
    \stepcounter{affN}%
    \csxdef{aff@num@#1}{\the\value{affN}}%
    \g@addto@macro\affblock{\textsuperscript{\csuse{aff@num@#1}}\csuse{aff@txt@#1}.\space}%
  \fi
  \csuse{aff@num@#1}%
}
\newcommand{\auth}[2]{\mbox{#1\textsuperscript{\aff@emit{#2}}}}
\newcommand{\aff@emit}[1]{\let\aff@sep\@empty\forcsvlist{\aff@one}{#1}}
\newcommand{\aff@one}[1]{\aff@sep\aff@get{#1}\let\aff@sep\aff@comma}
\newcommand{\aff@comma}{,}
\newcommand{\printaffils}{\affblock}
\makeatother

\newcommand\arcsec{$''$}%
\newcommand\arcmin{$'$}%

\usepackage{graphicx}
\usepackage{booktabs}
\usepackage{caption}
\usepackage[numbers,sort&compress,square]{natbib}
\usepackage{hyperref}
\hypersetup{
    colorlinks=true,
    linkcolor=RoyalNavy,
    citecolor=RoyalNavy,
    urlcolor=RoyalNavy
}

\usepackage[explicit]{titlesec}

\titleformat{\section}
  {\LARGE\bfseries\scshape\color{RoyalNavy}}
  {\thesection.}
  {0.6em}
  {#1}

\titleformat{\subsection}
  {\large\bfseries\scshape\color{RoyalNavy}}
  {\thesubsection.}
  {0.6em}
  {#1}

\titleformat{\subsubsection}
  {\normalsize\bfseries\color{RoyalNavy}}
  {\thesubsubsection.}
  {0.6em}
  {#1}

\titlespacing*{\section}{0pt}{2.6em}{0.0em}
\titlespacing*{\subsection}{0pt}{1.6em}{0.7em}
\titlespacing*{\subsubsection}{0pt}{1.2em}{0.5em}

\newcommand{\sectionrule}{%
  \vspace{-0.2em}
  \noindent\textcolor{RoyalNavy}{\rule{\textwidth}{0.8pt}}
  \vspace{0.2em}
}

\usepackage{fancyhdr}

\begin{document}


\includepdf[pages={1}]{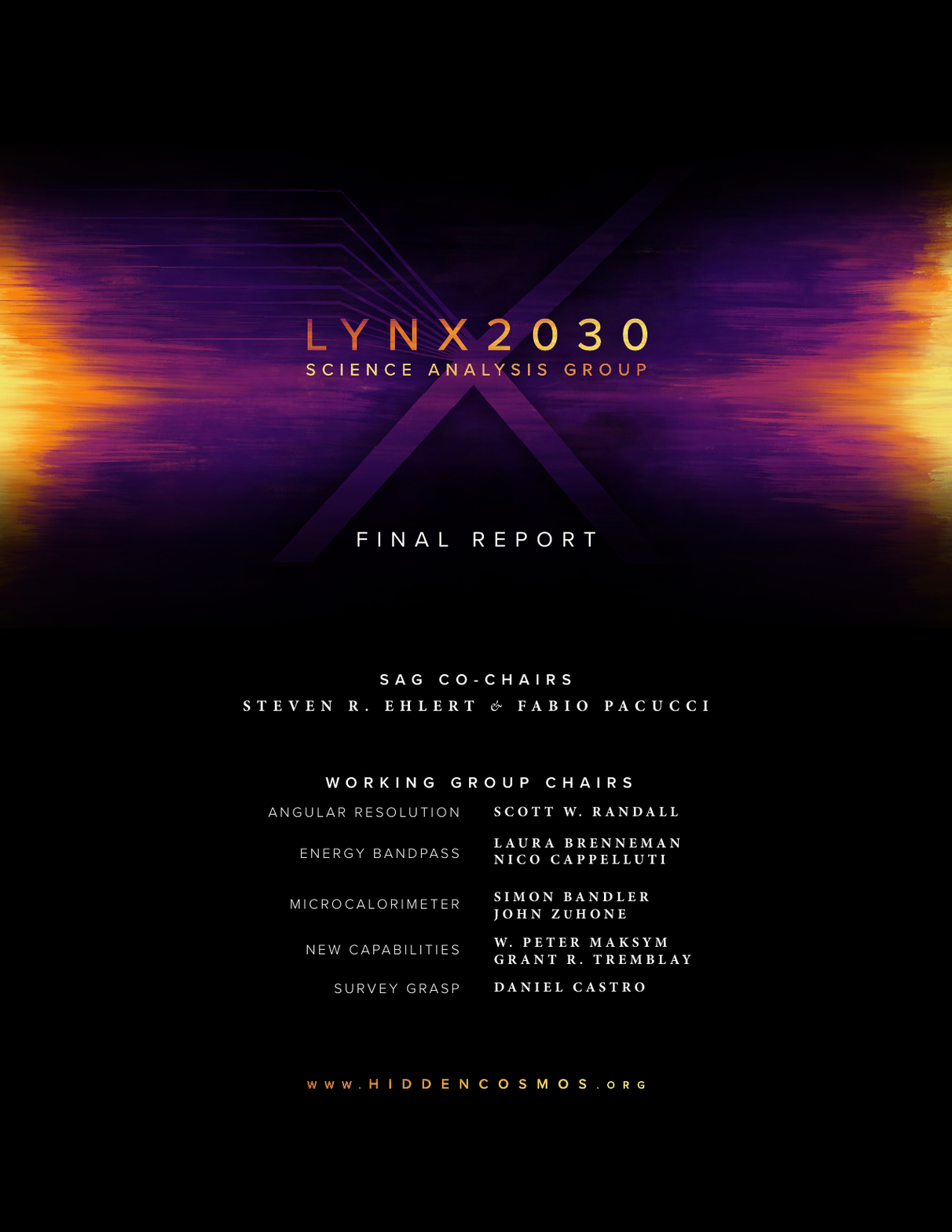}

\includepdf[pages={1}]{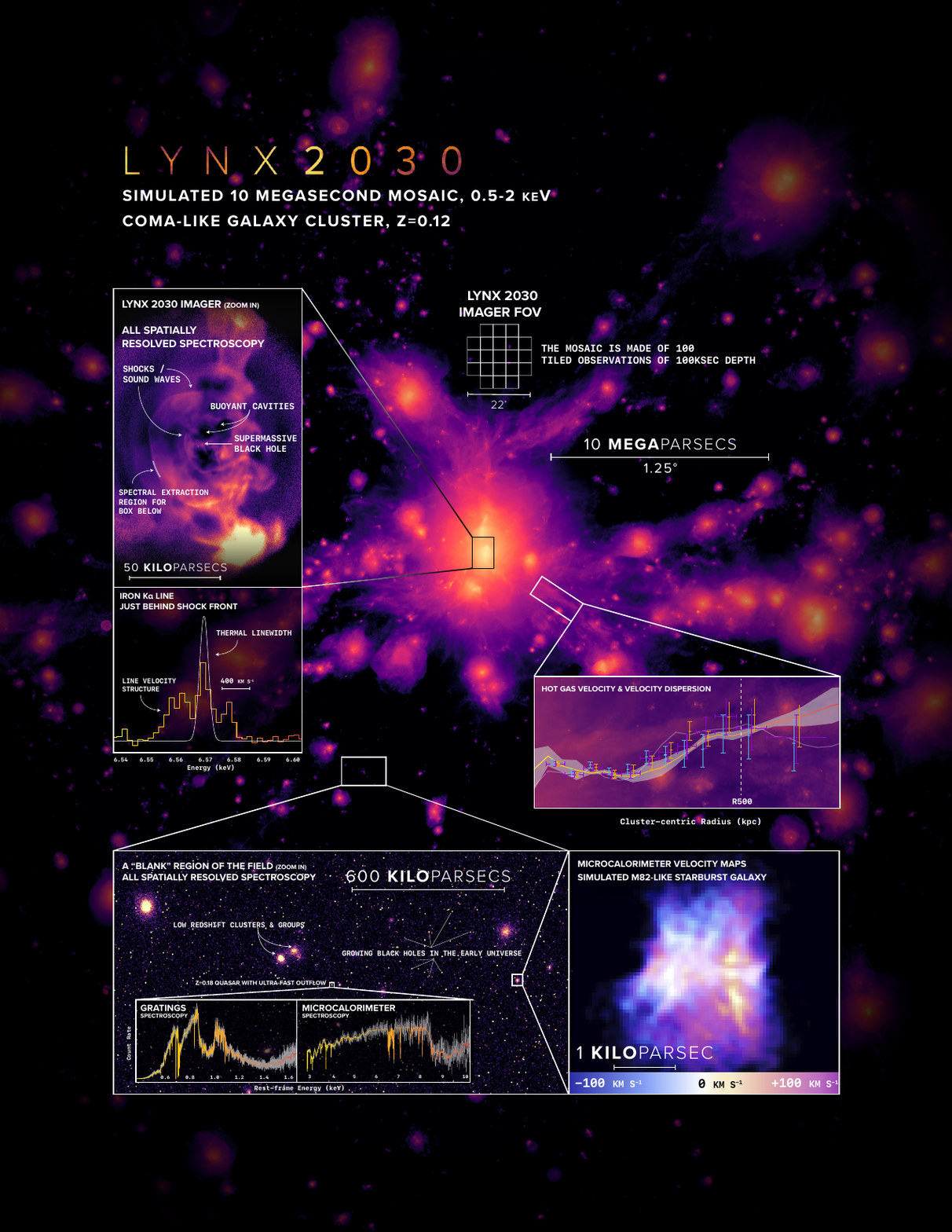}

\newpage
\thispagestyle{empty}

\vspace*{\fill}

\noindent\textcolor{RoyalNavy}{\rule{\textwidth}{1.2pt}}

\begin{abstract}

\begin{center}
{\huge\bfseries\color{RoyalNavy} Abstract}
\end{center}

\vspace{0.5cm}
{\large
\noindent
The Lynx2030 Science Analysis Group (SAG) was convened to reassess the scientific goals and technical drivers of the Lynx mission concept amid a rapidly evolving astrophysics landscape. Building on the original Lynx Concept Study, the SAG examined how recent discoveries, emerging facilities, and advances in instrumentation influence the scientific opportunities for a next-generation flagship X-ray observatory. Through focused working groups, the SAG investigated the scientific impact of enhanced capabilities: (i) improved angular resolution, (ii) broader bandpass coverage, (iii) an enhanced microcalorimeter, (iv) new capabilities and observing modes, and (v) larger fields of view. Across a broad range of topics, from the formation of the first black holes and the evolution of galaxies to the baryon cycle, compact objects, stellar explosions, multi-messenger astrophysics, and the dynamic high-energy Universe, the SAG finds that the scientific motivation for a Lynx-class observatory remains compelling and, in many areas, has significantly strengthened over the past decade, prominently through JWST's discovery of the ``Little Red Dots'', likely massive accreting black holes in infant galaxies whose nature is fundamentally an X-ray question. This report shows that modest extensions beyond the original Lynx design reference mission can unlock transformative science while preserving the observatory's core architecture. Powerful current and future facilities such as Roman, Rubin, JWST, SKA, ngVLA, LISA, and NewAthena highlight the unique role a high-angular-resolution, high-throughput X-ray observatory would play in the multi-wavelength and multi-messenger ecosystem of the 2030s and beyond. The findings of the Lynx2030 SAG confirm Lynx's central vision: an unprecedented view of the hot and energetic Universe, enabling discoveries that will define high-energy astrophysics in the coming decades.
}
\end{abstract}

\vspace{1em}
\noindent\textcolor{RoyalNavy}{\rule{\textwidth}{1.2pt}}

\vspace*{\fill}

\newpage
\thispagestyle{empty}

\begin{center}
{\Large\color{RoyalNavy}\scshape Lynx2030 Science Analysis Group}\\[0.3em]
{\color{RoyalNavy}\rule{0.45\textwidth}{0.8pt}}

\vspace{1.1cm}

{\normalsize\setstretch{1.25}

\definst{cfa}{Center for Astrophysics $\vert$ Harvard \& Smithsonian, 60 Garden St., Cambridge, MA 02138, USA}
\definst{sao}{Smithsonian Astrophysical Observatory, Center for Astrophysics $\vert$ Harvard \& Smithsonian, 60 Garden St., Cambridge, MA 02138, USA}
\definst{GSFC}{NASA Goddard Space Flight Center, Greenbelt, MD 20771, USA}
\definst{MSFC}{NASA Marshall Space Flight Center, Huntsville, AL 35812, USA}
\definst{umiami}{Department of Physics, University of Miami, Coral Gables, FL 33124, USA}
\definst{stanford}{Kavli Institute for Particle Astrophysics and Cosmology, Stanford University, 452 Lomita Mall, Stanford, CA 94305, USA}
\definst{SanAntonio}{Department of Physics and Astronomy, Trinity University, San Antonio, TX, USA}
\definst{Eureka}{Eureka Scientific, Inc., Oakland, CA, USA}
\definst{pmo}{Purple Mountain Observatory, Chinese Academy of Sciences, 10 Yuanhua Road, Nanjing 210023, People’s Republic of China}
\definst{asu}{Department of Physics, Arizona State University, Tempe, AZ 85287, USA}
\definst{um}{Department of Physics \& Astronomy, University of Manitoba, Winnipeg, Manitoba R3T 2N2, Canada}
\definst{cita}{CITA, University of Toronto, 60 St. George Street, Toronto, Ontario M5S 3H8, Canada}
\definst{msrc}{The Mallick Space Research Center, Hooghly, West Bengal 712701, India}
\definst{osu}{Department of Astronomy, The Ohio State University, 140 W 18th Ave, Columbus, OH 43210, USA}
\definst{ccapp}{Center for Cosmology and AstroParticle Physics, The Ohio State University, 191~W~Woodruff Ave, Columbus, OH 43210, USA}
\definst{ala}{Department of Physics \& Astronomy, University of Alabama, Tuscaloosa, AL 35487, USA}
\definst{brockport}{Department of Physics, SUNY Brockport, Brockport, NY 14420, USA}
\definst{notts}{School of Physics \& Astronomy, University of Nottingham, University Park, Nottingham NG7 2RD, UK}
\definst{umich}{Department of Astronomy, University of Michigan, Ann Arbor, MI 48109, USA}
\definst{iowa}{Department of Physics and Astronomy, The University of Iowa, Iowa City, Iowa 52242, USA}
\definst{msu}{Department of Physics and Astronomy, Michigan State University, East Lansing, MI 48824, USA}
\definst{MIT}{MIT Kavli Institute for Astrophysics and Space Research, Massachusetts Institute of Technology, Cambridge, MA 02139, USA}
\definst{NRAO}{National Radio Astronomy Observatory, 1180 Boxwood Estate Road, Charlottesville, VA 22903, USA}
\definst{columbia}{Columbia Astrophysics Laboratory and Department of Astronomy, Columbia University, 550 West 120th St., New York, NY 10027, USA}
\definst{chicago}{Department of Astronomy and Astrophysics, The University of Chicago, Chicago, IL 60637, USA}
\definst{IAPS}{Istituto di Astrofisica e Planetologia Spaziali, Via del Fosso del Cavaliere 100, 00133, Roma, Italy}


{\bfseries\Large On behalf of the Lynx2030 Science Analysis Group}

\vspace{0.9em}

{\Large\scshape\color{RoyalNavy}SAG and Working Group Chairs}

\vspace{0.35em}

\auth{Simon R. Bandler}{GSFC},
\auth{Laura W. Brenneman}{cfa},
\auth{Nico Cappelluti}{umiami},
\auth{Daniel Castro}{cfa},
\auth{Steven R. Ehlert}{MSFC}\textsuperscript{,\,$\star$},
\auth{W. Peter Maksym}{MSFC},
\auth{Fabio Pacucci}{cfa}\textsuperscript{,\,$\star$},
\auth{Scott W. Randall}{cfa},
\auth{Grant R. Tremblay}{cfa},
\auth{John ZuHone}{cfa}

\vspace{1.0em}

{\Large\scshape\color{RoyalNavy}Contributors}

\vspace{0.35em}

\auth{Steven W. Allen}{stanford},
\auth{Antara R. Basu-Zych}{GSFC},                     
\auth{Akos Bogdan}{cfa},                       
\auth{Joel N. Bregman}{umich},                       
\auth{Tamta Burduli}{asu},                      
\auth{Thomas Connor}{cfa},                        
\auth{Sanskriti Das}{stanford},                 
\auth{Casey DeRoo}{iowa},                        
\auth{Stephen DiKerby}{msu},                       
\auth{Paul A. Draghis}{MIT},                    
\auth{Martin Elvis}{cfa},
\auth{Giuseppina Fabbiano}{cfa},
\auth{Ralf K. Heilmann}{MIT},                      
\auth{Jimmy A. Irwin}{ala},                         
\auth{Amruta Jaodand}{cfa},                       
\auth{Margarita Karovska}{cfa},                 
\auth{Vinay L. Kashyap}{cfa},                       
\auth{Anthony A. Kerr}{NRAO},                          
\auth{Caroline Kilbourne}{GSFC},                     
\auth{Ralph Kraft}{cfa},                         
\auth{Jiangtao Li}{pmo},                            
\auth{Labani Mallick}{um, cita, msrc},                       
\auth{Herman L. Marshall}{MIT},                      
\auth{Michael L. McCollough}{cfa},                    
\auth{Anna Ogorza\l{}ek}{GSFC},                  
\auth{Frederik Paerels}{columbia},                       
\auth{Daniel Patnaude}{cfa},                      
\auth{Paul Plucinsky}{sao},
\auth{David Pooley}{SanAntonio, Eureka},
\auth{Frederick S. Porter}{GSFC},                        
\auth{Daniele Rogantini}{chicago},                     
\auth{Roger Romani}{stanford},                        
\auth{Helen R. Russell}{notts},                       
\auth{Kazuhiro Sakai}{GSFC},                         
\auth{Mark Schattenburg}{MIT},                  
\auth{Dan A. Schwartz}{cfa},
\auth{Malgorzata Sobolewska}{cfa},
\auth{Paolo Soffitta}{IAPS},                      
\auth{Alexey Vikhlinin}{cfa},
\auth{Daniel R. Wilkins}{osu,ccapp},                       
\auth{Scott Wolk}{cfa},                          
\auth{Ka-Wah Wong}{brockport},                          
\auth{Irina Zhuravleva}{chicago}                     

\par}
\end{center}

\vfill

{\small\setstretch{1.02}\noindent
{\scshape\color{RoyalNavy}Affiliations}\\[0.35em]
\printaffils\par
\vspace{0.5em}
\noindent$\star$\,Corresponding authors: Fabio Pacucci, Steven R. Ehlert.
\par}

\newpage
\thispagestyle{empty}
\section*{Executive Summary}
\sectionrule

The Lynx X-ray Observatory concept presented to the Astro2020 Decadal Survey \cite{LynxCSRFinal} defined a flagship-class, soft X-ray telescope built around three scientific pillars: black hole seeds at cosmic dawn, the drivers of galaxy evolution, and the energetic side of stellar evolution. To achieve these goals, such a mission requires a PSF comparable to that of \textit{Chandra}, coupled with $\sim 100 \times$ its effective area and a larger field of view (FoV). Pursuit of the science goals associated with these pillars was enabled by a subarcsecond-class optic with $\sim 2 \, \rm m^2$ of effective area at $1$ keV, a high-density imaging detector, an ultra-high-resolution microcalorimeter array, and a dispersive grating spectrometer. The High-Definition X-ray Imager (HDXI) included an array of 0.3\arcsec\ pixels with a $22' \times 22'$ FoV; the Lynx X-ray Microcalorimeter (LXM) was designed with a main array of $5' \times 5'$ FoV (1\arcsec\ pixels) and an inner enhanced main array of $1' \times 1'$ (0.5\arcsec\ pixels).

In the half-decade since Astro2020, the Lynx science case has, in many ways, been strengthened. Most prominently, JWST's discovery of the ``Little Red Dots'' (LRDs): extremely compact, red-continuum sources with broad Balmer lines, now widely interpreted as massive, X-ray-weak accreting black holes in the hearts of infant galaxies. This discovery has sharpened the demand for the deep, high-resolution X-ray imaging spectroscopy that only a Lynx-class facility can deliver. Whether their X-ray weakness reflects intrinsic super-Eddington accretion, Compton-thick obscuration, or reprocessing in dense ionized cocoons, the question is fundamentally an X-ray one: distinguishing these scenarios for individual sources at $z \gtrsim 5$ requires the combination of large effective area, arcsecond imaging to beat down confusion and background, and the spectral resolution to characterize absorption and ionized winds—precisely the parameter space Lynx was designed to own. More broadly, recent discoveries from IXPE, XRISM, and JWST have underscored the necessity of capabilities such as high-resolution X-ray polarimetry and rapid-readout timing, while reaffirming the centrality of the optic and the microcalorimeter.

The key enabling technologies of the Lynx design have meanwhile advanced in their Technology Readiness Level (TRL), driven largely by six years of continued technology development partly coupled to a cohort of Probe- and Explorer-class studies: STAR-X and AXIS (high-angular-resolution imaging enabled by silicon meta-shell optics), LEM (microcalorimeters), Arcus (critical-angle transmission gratings), and HEX-P (hard X-ray response), together with the on-orbit successes of IXPE and XRISM. X-ray microcalorimeters remain the unequivocal priority for high-energy-resolution imaging spectroscopy: XRISM/Resolve has flight-validated the core measurement, and the LEM Probe study has advanced the large-format, fine-resolution arrays directly relevant to the LXM, with the technology advancing robustly toward TRL-5. The net assessment is encouraging: the two technologies that most distinguished Lynx from its predecessors (i.e., its arcsecond, square-meter optic and the $>100{,}000$-pixel microcalorimeter) have both advanced materially since the previous development cycle, while the scalable integration of thousands of mirror segments into aligned meta-shells remains the central technical and programmatic risk.

Against this backdrop, the Lynx2030 SAG convened five working groups (Angular Resolution, Energy Bandpass, Microcalorimeter, New Capabilities, and Survey Grasp), taking the Lynx2020 Concept Study Report (CSR) as a foundational baseline, and asking what is newly enabled by enhanced capabilities: improved angular resolution, broader bandpass coverage, an enhanced microcalorimeter, new capabilities and observing modes, and larger fields of view, which were not part of the 2020 baseline but are now sufficiently mature to be considered for mating with a Lynx-class optic. In addition, new launch and spacecraft capabilities may open doors beyond the original Lynx concept. The case for sustained, front-loaded investment in these enabling technologies is, accordingly, stronger today than it was in 2020.

\clearpage
\thispagestyle{empty}
\begin{center}
\vspace*{-0.5em}
{\Large\bfseries\scshape\color{RoyalNavy}\textls[60]{Contents}}\\[-0.7em]
\customsepnavy
\end{center}
\vspace{-0.5em}
{\makeatletter\@starttoc{toc}\makeatother}
\clearpage

\newpage
\thispagestyle{empty}
\section*{List of Acronyms}

\vspace{1cm}

\begin{center}
\footnotesize
\renewcommand{\arraystretch}{1.15}
\setlength{\tabcolsep}{4pt}

\begin{longtable}{|C{2.2cm}|L{9.4cm}|}

\hline
\rowcolor[HTML]{203864}
\color{white}\textbf{Acronym} &
\color{white}\textbf{Definition} \\
\hline
\endfirsthead

\hline
\rowcolor[HTML]{203864}
\color{white}\textbf{Acronym} &
\color{white}\textbf{Definition} \\
\hline
\endhead

\textbf{ADD} & Advancement Degree of Difficulty \\
\hline
\textbf{ADR} & Adiabatic Demagnetization Refrigerator \\
\hline
\textbf{ASIC} & Application-Specific Integrated Circuit \\
\hline
\textbf{AXIS} & Advanced X-ray Imaging Satellite \\
\hline
\textbf{CAT} & Critical-Angle Transmission (gratings) \\
\hline
\textbf{CATXGS} & CAT X-ray Grating Spectrometer \\
\hline
\textbf{CCD} & Charge-Coupled Device \\
\hline
\textbf{CMOS} & Complementary Metal-Oxide-Semiconductor \\
\hline
\textbf{CSR} & Concept Study Report \\
\hline
\textbf{EMA} & Enhanced Main Array \\
\hline
\textbf{eXTP} & enhanced X-ray Timing and Polarimetry mission \\
\hline
\textbf{FoV} & Field of View \\
\hline
\textbf{GPD} & Gas Pixel Detector \\
\hline
\textbf{GSPS} & Giga-Samples Per Second \\
\hline
\textbf{HDXI} & High-Definition X-ray Imager \\
\hline
\textbf{HEMT} & High-Electron-Mobility Transistor \\
\hline
\textbf{HEX-P} & High-Energy X-ray Probe \\
\hline
\textbf{HPD} & Half-Power Diameter \\
\hline
\textbf{HWO} & Habitable Worlds Observatory \\
\hline
\textbf{IFU} & Integral Field Unit \\
\hline
\textbf{IXPE} & Imaging X-ray Polarimetry Explorer \\
\hline
\textbf{LEM} & Line Emission Mapper \\
\hline
\textbf{LXM} & Lynx X-ray Microcalorimeter \\
\hline
\textbf{MDP} & Minimum Detectable Polarization \\
\hline
\textbf{MRL} & Manufacturing Readiness Level \\
\hline
\textbf{\boldmath$\mu$MUX} & Microwave Multiplexing (readout) \\
\hline
\textbf{OBF} & Optical Blocking Filter \\
\hline

\textbf{PSF} & Point-Spread Function \\
\hline

\textbf{SMO} & Silicon Meta-shell Optics \\
\hline
\textbf{SQUID} & Superconducting Quantum Interference Device \\
\hline
\textbf{SRXO} & Segmented Replicated X-ray Optics \\
\hline
\textbf{STAR-X} & Survey and Time-domain Astrophysical Research eXplorer \\
\hline
\textbf{TES} & Transition-Edge Sensor \\
\hline
\textbf{TRL} & Technology Readiness Level \\
\hline
\textbf{UHR} & Ultra-High-Resolution (sub-array) \\
\hline
\textbf{ULE} & Ultra-Low-Expansion (glass) \\
\hline
\textbf{UVEX} & UltraViolet Explorer \\
\hline
\textbf{XRISM} & X-ray Imaging and Spectroscopy Mission \\
\hline

\end{longtable}
\end{center}

\clearpage

\newpage
\pagenumbering{arabic}

\section{Angular Resolution Working Group}
\sectionrule

This section represents the work of the Lynx2030 SAG Angular Resolution working group. The goal is to update science cases for a Lynx-like mission in the 2030s and beyond. ``Lynx2020,'' as described in the CSR for Astro2020, is taken as a given baseline. Here, we focus on new and updated science cases enabled by new technologies, observatories, and scientific discoveries, with a focus on what can be achieved with an angular resolution less than the 0.5\arcsec\ half-power diameter (HPD) proposed for Lynx2020. We assume a lower limit of $\sim0.1$\arcsec\ HPD, since we conclude that finer angular resolution will very likely require new technologies that are currently at a lower TRL (e.g., X-ray interferometry).

We identify three pillar science cases that are fundamentally transformed by $\sim0.1''$ imaging: measuring AGN feedback out to and across cosmic noon (\S\ref{sec:feedback_morph}, \S\ref{sec:feedback_rad}), understanding early black hole seeding and growth (\S\ref{sec:blackholes}), and resolving hot gas inside the Bondi radius of nearby super-massive black holes (SMBHs, \S\ref{sec:bondi}). We also identify several non-pillar science cases that benefit significantly from angular resolutions of $\lesssim 0.5''$. All science cases and related requirements are summarized in Table~\ref{tab:science_summary}.

\subsection{Morphological Signatures of AGN Feedback at Cosmic Noon}
\label{sec:feedback_morph}

Galaxy formation is dominated by the cosmic struggle between gravity and feedback.  Gravity pulls mass together to form stars and galaxies, like our own Milky Way. However, if our models only include gravity, galaxies grow far too big and form far too many young stars. Additional physics is needed to explain the galaxies we see around us. The energetic processes around black holes, known collectively as feedback, are now thought to provide the missing counterbalance to gravity, driving gas back out of massive galaxies and regulating their growth over cosmic time \citep{DiMatteo05,Croton06,Vogelsberger14}.
Feedback also has important consequences for galaxy evolution, the growth of structure, and the distribution of metals over cosmic time, and must be included in state-of-the-art cosmological simulations to produce results that resemble the observed Universe. 

The clearest evidence for black hole feedback is found in the most massive galaxies, which are surrounded and pervaded by hot atmospheres at millions of degrees \citep{McNamara12}. Images from the \textit{Chandra} X-ray observatory reveal vast cavities where jets launched by the central black hole have displaced the hot gas by inflating large bubbles filled with radio-emitting plasma \citep{McNamara00,FabianPer00,Churazov02,Li22, Li24}.  Our current understanding of radio-mode feedback is almost entirely shaped by deep \textit{Chandra} observations of the nearest, most massive central cluster galaxies.  Although they provide crucial insights, these systems are unlikely to represent the wider galaxy population, particularly the individual Milky Way-sized galaxies that contain most of the stars in the Universe, or the height of AGN and stellar activity at cosmic noon.  With sharp imaging and a dramatic increase in sensitivity, a Lynx-like mission will extend our observations of feedback from the behemoths to span the full range of hot atmospheres around massive galaxies, and back to the critical peak of galaxy formation at cosmic noon, $\sim10$ billion years ago.

\begin{figure}[H]
\begin{minipage}{\textwidth}
\centering
\includegraphics[width=0.32\columnwidth]{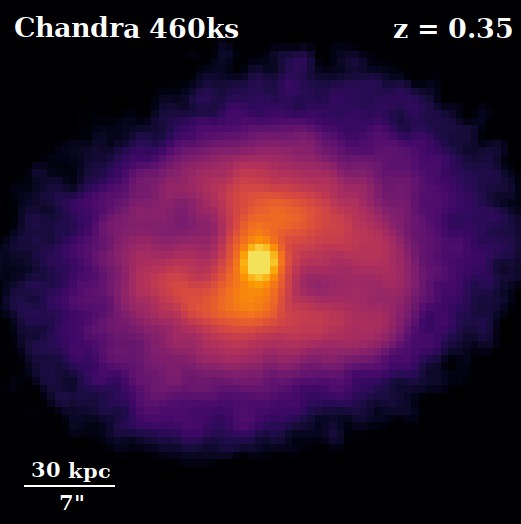}
\includegraphics[width=0.32\columnwidth]{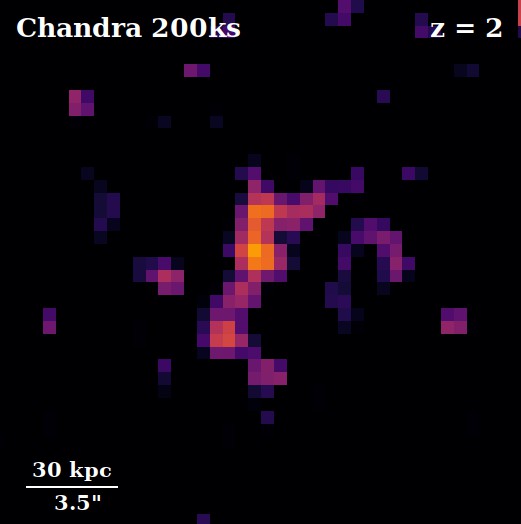}
\includegraphics[width=0.32\columnwidth]{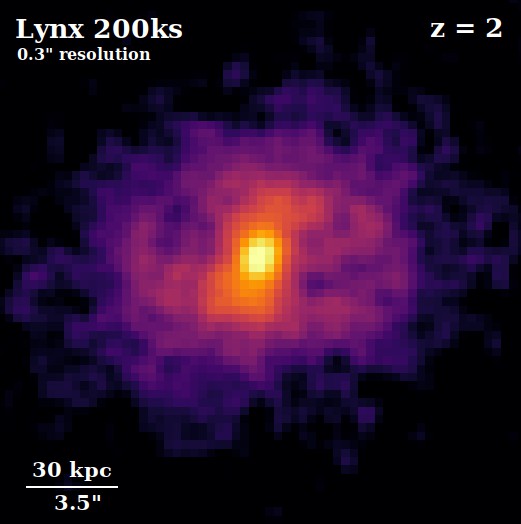}
\caption{\textbf{Left:} \textit{Chandra} X-ray image of RBS797 at $z=0.35$ showing a bright nucleus and two X-ray cavities, each 30\,kpc in size, in the E-W direction \citep{Schindler01,Ubertosi21}.  \textbf{Center:} A simulated \textit{Chandra} image of this cluster at $z=2$.  The cluster has been scaled to $2\times10^{14}$\,M$_{\odot}$, reduced in angular size, cosmologically dimmed etc.  The cavities are not detected.  \textbf{Right:} A simulated \textit{Lynx} image at 0.3\arcsec\ resolution, with clearly visible cavities at $z=2$.  Images have been smoothed with a 2D Gaussian with $\sigma=2$\,pixels.  Based on an original figure by M. McDonald.}
\label{fig:cavities}
\end{minipage}
\end{figure}

\begin{figure}[H]
\begin{minipage}{\textwidth}
\centering
\includegraphics[width=0.24\columnwidth]{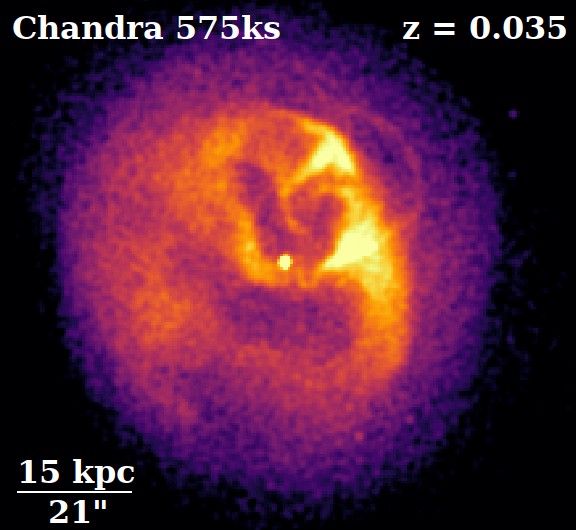}
\includegraphics[width=0.24\columnwidth]{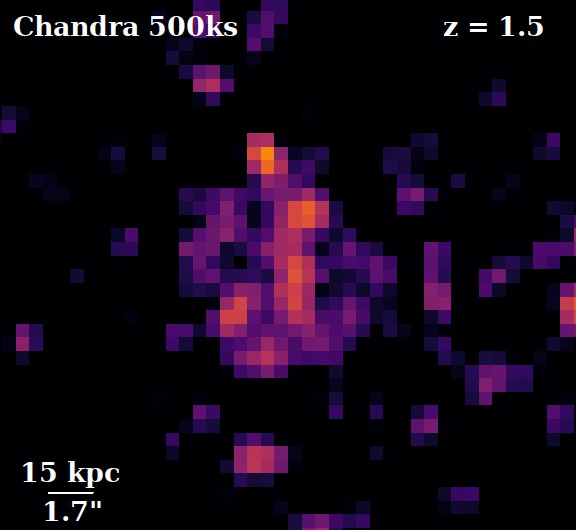}
\includegraphics[width=0.24\columnwidth]{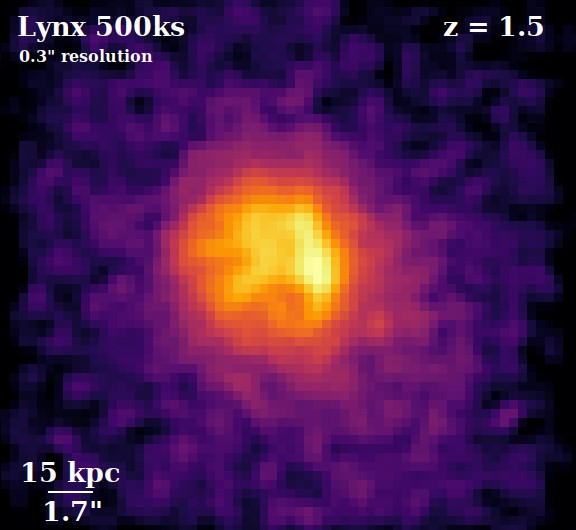}
\includegraphics[width=0.24\columnwidth]{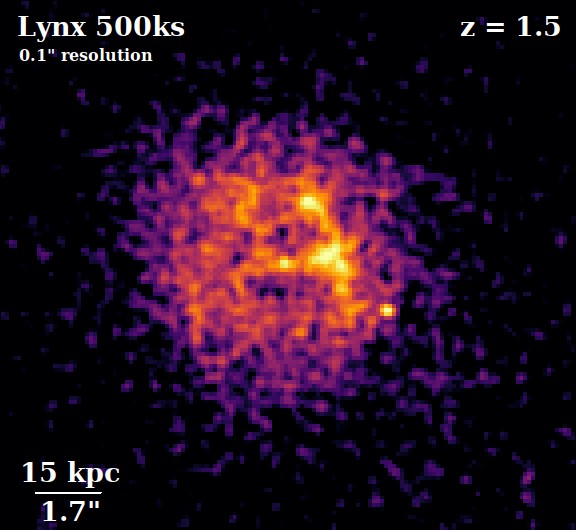}
\caption{Left: \textit{Chandra} X-ray image of A2052 at $z=0.035$ showing a bright nucleus and two X-ray cavities, each 15\,kpc across, in the N-S direction \citep{Blanton01,Blanton11}.  Left Center: A simulated \textit{Chandra} image of this cluster at $z=1.5$.  The cluster has been reduced in angular size, cosmologically dimmed etc.  The cavities are not detected.  Right Center: Simulated \textit{Lynx} images with 0.3\,arcsec and 0.1\,arcsec (right) spatial resolution.  These more typically sized cavities are detected at $z=1.5$ with the increased spatial resolution.}
\label{fig:A2052cavities}
\end{minipage}
\end{figure}

The energy output of an AGN can be coupled to the surrounding diffuse gas via cavity inflation, shocks, and outflows. Existing and upcoming observatories (e.g., ALMA, JWST, HST, VLBA, Roman, GMT, SKA) can map colder gas and stellar populations at the relevant physical scales, with sub-arcsecond resolution, at cosmic noon. However, X-ray observations are required to characterize the hot gas and its coupling to AGN activity, as observed in the local Universe \citep{Blanton11, 2015ApJ...805..112R, Li22}.
The relevant physical scale for cavities and shocks in galaxies and groups is a few to tens of kpcs. Therefore, $\sim 1$~kpc resolution is needed to characterize these features in the X-ray. 1~kpc corresponds to $\sim$0.5\arcsec at $z\sim0.1$, limiting what can be done with {\it Chandra} to the relatively local Universe. However, at cosmic noon ($z\sim1-2$), 1~kpc~$\approx$~0.1\arcsec. Therefore, an X-ray imager with $\sim$~0.1\arcsec\ angular resolution can effectively probe signatures of AGN feedback in the hot diffuse gas during the critical era of galaxy formation known as cosmic noon (Figures~\ref{fig:cavities}~and~\ref{fig:A2052cavities}).

This science would benefit from maximal on-axis spatial resolution, which is the limiting factor when trying to resolve cavities in targeted observations of high-redshift clusters.  High-redshift targets will be selected from upcoming Sunyaev-Zel'dovich surveys from SPT, ACT, the Simons Observatory, and CMB-S4, which will provide robust catalogs of massive, virialized systems at $z>1.5$ \citep{Mantz19}.  There are clear synergies with next-generation radio observatories, such as SKA and ngVLA, which will reveal the jet and lobe emission.  Multi-wavelength observations of cool gas nebulae, with e.g. ALMA, JWST, and ELTs, will reveal the triggers of thermal instability and the impact on cool clouds and star formation.

Observations \citep{Panagoulia14} and simulations (e.g. Figure~\ref{fig:cavities} and Figure~\ref{fig:A2052cavities}) of X-ray cavities in massive galaxies and clusters have demonstrated that unambiguous detections require $>20,000$\,cts within a radius of 10\, kpc or, similarly, $>10$\,cts/pix for the smallest cavities that can be resolved.  For a sample of low-mass early-type galaxies, comparable in size to the Milky Way (e.g. ATLAS$^{\mathrm{3D}}$, \cite{Cappellari11}), this could be achieved in typical exposure times of $5-50\,$ks.  For a high-redshift cluster sample selected from SZ surveys, and assuming cavities with comparable morphologies, exposure times of $100-500\,$ks would be required.

\subsection{SMBH -- Host Galaxy Interaction: Radiative Feedback}
\label{sec:feedback_rad}

It is a well-accepted paradigm in astrophysics that galaxy and nuclear supermassive black hole (SMBH) formation and evolution are linked ($M$--$\sigma$), and that accreting SMBHs, i.e., active galactic nuclei (AGNs), could shape this evolution via the photons produced in the accretion disk and corona (radiative feedback), and through winds and jets (kinetic feedback). The \textit{Chandra X-ray Observatory} has observed efficient feedback in radio-loud cluster galaxies, but most AGNs are not radio-loud. For these systems, while there is consensus that the AGN may be a source of feedback, conclusive observational proof is still lacking. Is the AGN power--ISM coupling effective, and which of winds, jets, and radiation is the dominant feedback mechanism? As shown by long \textit{Chandra} observations of nearby AGNs at $\sim 0.2^{\prime\prime}$ resolution \citep{Maksym17, Fabbiano18,Li22}, each of these mechanisms has X-ray signatures in the $\sim 0.3$--$7.0~\mathrm{keV}$ range, complementing high-resolution observations with \textit{HST}, \textit{JWST}, ALMA,  VLA, Euclid, Roman, and others. To constrain the physical mechanisms at play, a future X-ray flagship will need to have \textit{Chandra's} sub-arcsecond resolution, or better, $\sim 0.1^{\prime\prime}$, 
to provide spatial resolution comparable to that of existing and forthcoming multiwavelength observatories, combined with spatially resolved high-resolution spectroscopy, e.g., an X-ray calorimeter, and 100 times larger collecting area to achieve reasonable S/N in many spatial/spectral resolution elements. This observatory will allow in-depth studies in the local universe and extend the study volume by a factor of $\sim 1000$. It will also enable the detailed study of the more extended ionization regions of more powerful quasars out to $z \sim 0.5$.



ESO~428-G014 provides a representative scaling for spatially resolved studies of feedback in nearby, radio-quiet AGN. The
existing analysis is based on a coadded 215-ks \textit{Chandra}/ACIS-S
observation at a distance of approximately 40~Mpc, corresponding to
184~pc per arcsecond \citep{Fabbiano18}.
Assuming an observatory with an effective area 100 times that of \textit{Chandra}, a 20 ks exposure of ESO 428-G014 itself would yield ~46,000 counts from a central region of 1.5\arcsec\ radius, in the 0.3-7 keV band. With ACIS ($\sim$~150~eV spectral resolution) some emission lines are detected with $\sim$90~cts per resolution element. With ten times the counts and $\sim$10~eV spectral resolution, we should be able to see a rich emission line spectrum (with $\sim65$ cts/resolution element, i.e., $\sim12\%$ line flux uncertainties). From the 8\arcsec\ ($\sim$1.5~kpc) extended ionization cone, we expect only a factor of two fewer counts, $\sim$23,000~cts, so similar considerations apply for higher resolution spectra. Spectral binning would lead to spectral characterization/modeling of several emission regions. AGNs at smaller distances will yield higher S/N data. From the point of view of spatial analysis, assuming a 0.3\arcsec\ resolution element, we would get $\sim$2200 resolution elements in the extended area seen with \textit{Chandra}, yielding on average a 3$\sigma$ detection with $\sim10$~cts per pixel.

With a ~20 ks exposure (1/10 of that of \textit{Chandra}), we would get data comparable to that from the coadded \textit{Chandra} exposure at a distance of $D\sim120$ Mpc (i.e., out to the Coma Cluster). These short observations would allow the study of a few 100 similar galaxies that are currently not reachable with \textit{Chandra}. Of course, exposures of $\sim$100~ks will allow the study of similar AGNs out to $D \sim 270$~Mpc. 

In the $\sim 1000 \times$ larger volume, there are rarer but more luminous quasars that are too faint for \textit{Chandra}, but relatively easy targets for a Lynx-like observatory. These Type II Quasars ($z \sim 0.1-0.4$, $\sim 400-2000$~Mpc luminosity distance) have galaxy-size extended ionization regions several arcseconds in radius \citep{2018ApJ...868...14S}. We can estimate results from a Lynx-like mission using the photoionization [O~III]/X-ray ratio of $\sim$10 (from the \cite{2009ApJ...704.1195W} HRI study of NGC 4151). To reach science goals comparable to those of the \textit{Chandra} observation of ESO~428-G014, one would need a $\sim$200~ks exposure. However, with a 20~ks exposure, it would be easy to establish the presence of extended X-ray-emitting regions. Therefore, one could design a snapshot survey to identify candidates for further study, obtain sample statistics, and then follow up with deeper observations.

\subsection{Early Black Hole Seeding and Growth}
\label{sec:blackholes}

Understanding the origin of supermassive black holes (SMBHs) remains one of the major unsolved problems in astrophysics. Despite (and because of!) significant advances with JWST, the high-$z$ Universe still features profound unanswered questions. Competing seed formation scenarios predict distinct black hole masses, occupation fractions, and luminosity functions during the first billion years of cosmic history \citep{Inayoshi20,Volonteri21}. In particular, light seed models (i.e., with $M_\bullet \lesssim 10^3 M_\odot$) in which black holes originate from the remnants of Population III stars, predict a significantly different abundance of low-luminosity AGN at ($z \gtrsim 8$) than heavy seed models (i.e., with $ 10^4 M_\odot  \lesssim M_\bullet \lesssim 10^6 M_\odot$), in which black holes form through the direct collapse of primordial gas clouds \citep{Ricarte18,Cappelluti_2023_AXIS}.
Measuring the demographics of accreting black holes during the epoch of reionization, as well as attempting direct X-ray detection of heavy seeds, are the most straightforward ways to constrain the population of the first black holes formed in the Universe \citep{Pacucci_2022_search,Ricarte18,Cappelluti_2023_AXIS}.

X-ray observations provide a uniquely robust probe of this early population because they remain sensitive to obscured sources that are difficult to identify with optical and infrared diagnostics alone \citep[e.g.,][]{Goulding23,Bogdan24}. A next-generation X-ray flagship capable of detecting large samples of faint AGN at $z > 8$ would enable the first unbiased census of early black hole growth and constrain the nature of black hole seeds \citep{Cappelluti_2023_AXIS}.

High angular resolution is critical for maximizing the scientific return of deep X-ray surveys targeting the first accreting black holes. The earliest galaxies, as well as the newly discovered JWST's Little Red Dots, are compact systems with characteristic sizes of only a few hundred parsecs \citep{Matthee_2023}. At $z \sim 8-10$, an angular scale of 0.1\arcsec\ corresponds to approximately 0.5 kpc, while 0.5\arcsec\ corresponds to 2.5 kpc. Sub-arcsecond imaging therefore allows the nuclear X-ray source to be separated from nearby emission associated with, e.g., star formation, X-ray binaries, and neighboring galaxies (see, e.g., the case of the Little Red Dots) in dense environments, where early structure formation occurs \citep{Matthee_2023,Yue_2024_Xray}. 

Angular resolution also determines the confusion limit of ultra-deep surveys. Since the beam area scales as $A_{\rm beam} \propto \theta^2$, where $\theta$ is the PSF, improving it from 0.5\arcsec to 0.1\arcsec\ reduces the beam area by a factor of 25. This effect dramatically lowers the probability of source blending and increases the effective depth of deep surveys aimed at detecting the faintest accreting black holes. Furthermore, improved angular resolution enables the identification of dual black holes in merging systems, providing direct knowledge into the hierarchical assembly of the earliest galaxies and their central SMBHs \citep{Volonteri21}.

The science case relies strongly on synergies with JWST, Roman, and future extremely large telescopes. While these facilities will identify and characterize candidate host galaxies, X-ray observations provide the cleanest signature of SMBH accretion \citep{Goulding23,Bogdan24,Cappelluti_2023_AXIS}. Accurate source localization is, hence, critical. A 0.1\arcsec\ X-ray position can be matched directly to high-resolution optical and infrared imaging, minimizing counterpart ambiguities.

The science case also benefits from the fact that the observed-frame of $0.3-10$ keV corresponds to rest-frame energies of several to tens of keV at $z > 8$. This allows the detection of moderately obscured accretion that may be missed by other techniques. Consequently, high-angular-resolution X-ray surveys provide a more complete census of early black hole growth than can be obtained from optical or infrared surveys alone.

The primary measurement of such a science case is source detection and localization, rather than spatially resolved imaging. Hence, relatively modest photon statistics are sufficient for many of the key science goals. Approximately 10-20 net counts are generally adequate for strong source detection and positional association with a candidate host galaxy. Samples of hundreds of such faint sources can then be used to construct the high-$z$ AGN luminosity function and distinguish between competing seeding scenarios \citep{Ricarte18,Cappelluti_2023_AXIS}.

For brighter objects, 50-100 counts enable basic spectral characterization and obscuration estimates. However, because the principal science driver is detecting and identifying extremely faint AGN, the most important instrumental requirement is maintaining sub-arcsecond angular resolution at the flux limits where source confusion would otherwise dominate.

\subsection{Resolving Hot Gas Inside the Bondi Radius of SMBHs}
\label{sec:bondi}

\begin{figure}[h]
\centering
\includegraphics[width=1.00\textwidth]{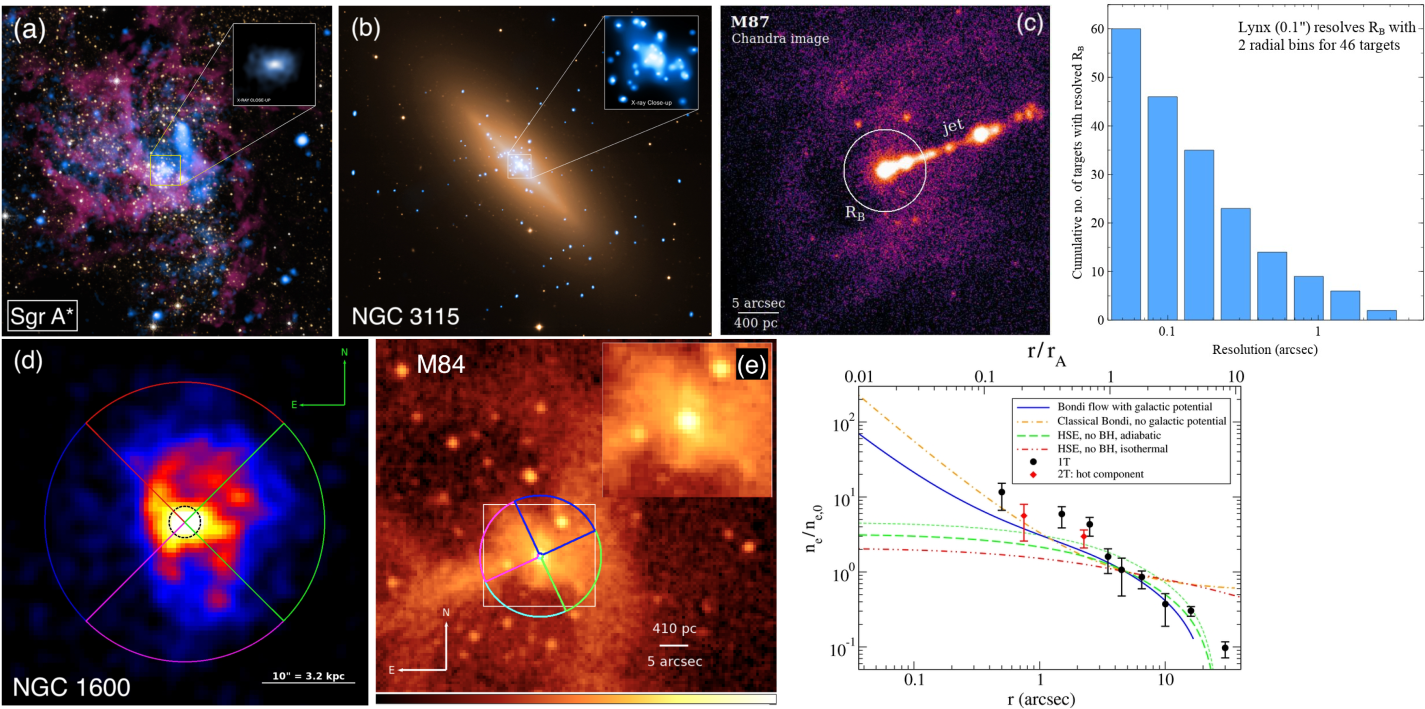}
\caption[Resolved Bondi regions with \textit{Chandra}]{\textbf{Left:} LLAGNs with Bondi regions spatially resolved with \textit{Chandra}: Sgr~A$^*$ \citep{Wan+13}, NGC~3115 \citep{WIY+11,WIS+14}, M87 \citep{RFM+15,Rus+18}, NGC~1600 \citep{RW21}, and M84 \citep{BRR+23}; adapted from \cite{2024Univ...10..278W}.  In the first two panels, X-ray emission is shown in blue, and is also visible in the insets (image credits: NASA/CXC/SAO).  The remaining panels are X-ray images. 
\textbf{Upper right:} Cumulative distribution of Bondi-radius sizes for a sample of SMBHs, illustrating the ability to resolve hot gas within $R_B$ for a significant number of sources with sub-arcsecond resolution. Chandra is unable to do much better than the five Bondi radii already studied.
\textbf{Lower right:} Density profiles predicted by different RIAF models and models without a black hole \citep{WIS+14}. Chandra can probe only the hot gas near the Bondi radius. Distinguishing between these models requires resolving the hot gas down to $\lesssim0.1$\arcsec scales.
}
\label{fig:bondi}
\end{figure}

The accretion of hot gas onto supermassive black holes (SMBHs) represents one of the most energetically consequential processes in the Universe, regulating both SMBH growth and the co-evolution of galaxies \citep{Fabian_2012,MN07}. Within the gravitational capture scale for the ambient hot gas — quantified by the Bondi radius, $R_B = 2G M_{\rm BH}/c_s^2$, where $c_s$ is the ambient sound speed — the dynamics of the infalling gas undergo a fundamental transition from the large-scale interstellar medium (ISM) to the accretion flow \citep{Bon52}. Probing this transition region is central to understanding how black holes feed and how they drive feedback into their host galaxies \citep{ADA+06, RME+13, Pillepich_2018, 2024Univ...10..278W, BM99,QN00,Bag+03,2010ApJ...710..755G,Blandford2022,Lalakos2022,Guo2023,Olivares2023}.

The five nearest SMBHs for which Chandra has spatially resolved the Bondi radius — Sgr A*, NGC 3115, M87, NGC 1600, and M84 (left images in Figure~\ref{fig:bondi}) — are all classified as low-luminosity AGNs \citep[LLAGNs;][]{Ho08}. These systems are believed to accrete in the hot, radiatively inefficient accretion flow (RIAF) mode, in contrast to the cold, thin-disk accretion that characterizes luminous quasars \citep{Yuan_Narayan_2014}. The RIAF regime is particularly poorly constrained observationally because the relevant physical scales have, until now, been beyond the reach of available instrumentation.

A key prediction of this hot RIAF theory is a steep rise in gas temperature within $R_B$, driven by adiabatic compression as gas falls into the potential well \citep{Yuan_Narayan_2014}. Contrary to this expectation, none of the five resolved systems exhibit such a temperature upturn \citep{Wan+13,WIY+11,WIS+14,RFM+15,Rus+18,RW21,BRR+23,2024Univ...10..278W}. Instead, the temperature structure within the Bondi region is complex and suggestive of a multiphase medium, with a cooler component at $kT \approx 0.2$--$0.3$\,keV embedded within the hotter ambient gas. Simultaneously, the measured density profiles within $R_B$ are surprisingly shallow, inconsistent with pure Bondi inflow or the classical advection-dominated accretion flow \citep[ADAF;][]{Ich77,RBB+82,Narayan_1994}, indicative of powerful outflows that reduce the net mass accretion rate \citep[advection-dominated inflow-outflow solutions or ADIOS;][]{BB99,Beg12,Yuan_Narayan_2014}.

These results align qualitatively with more recent three-dimensional magnetohydrodynamic (MHD) simulations, which predict that large-scale accretion within $R_B$ can be highly chaotic: cooler gas may rain inward in some directions while hotter, outflowing gas dominates other solid angles \citep[see review by][]{Yuan_Narayan_2014}. However, the observational distinction between this chaotic cold accretion \citep[CCA;][]{PSB+17} scenario and alternative models — such as magnetically arrested disks \citep[MADs;][]{Igumenshchev_2003, Igumenshchev2008, McKinney2012, Tchekhovskoy2012} or convection-dominated accretion flows \citep[CDAFs][]{NYA00,QG00,AIQ+02} — hinges critically on the radial density and temperature structure at scales deep within $R_B$.

The fundamental limitation of existing Chandra observations is angular resolution. Even with unlimited exposure time, Chandra is unable to do much better than the five Bondi radii already studied, whereas a 0.1\arcsec\ resolution would increase the number of resolvable Bondi radii by nearly an order of magnitude (Figure~\ref{fig:bondi}). The Bondi radii of even the largest accessible SMBHs subtend angles of only a few arcseconds on the sky, permitting at most a handful of independent radial bins within $R_B$. The gas properties measured near $R_B$ are therefore still transitional — not yet in the regime where different accretion models diverge (lower right panel in Figure~\ref{fig:bondi}). 
Measuring the density slope and temperature anisotropy at $r \ll R_B$ is the crucial observational step required to break the degeneracy among competing theoretical frameworks.

A spatial resolution of $\lesssim 0.1$\arcsec\ (Lynx2030) would transform studies of gas within the Bondi radius by providing an approximately fivefold improvement over Chandra (lower right panel in Figure~\ref{fig:bondi}). Such resolution would enable measurements of the density profile with roughly ten radial bins inside $R_B$ for systems with $R_B > 1$\arcsec, reaching $r \lesssim 0.1 R_B$, where theoretical models make distinct predictions for the density slope (e.g., ADAF: $-1.5$; CDAF: $-0.5$; ADIOS: $-1$ to $-1.5$; MAD: $-0.6$ to $-1.1$; \cite{Yuan_Narayan_2014,Cho2025,Lalakos2025}). It would also enable hardness-ratio maps with a spatial resolution of $r \approx 0.25 R_B$, allowing us to test whether the thermal structure is anisotropic, as predicted by inflow–outflow models, or more randomly distributed, as expected in chaotic accretion. In addition, spatially resolved spectroscopy within $R_B$ would constrain the multiphase gas structure. Both diagnostics are achievable in 300--500\,ks with a Chandra-class effective area. With a tenfold increase in effective area relative to Chandra, full temperature maps of M87 could be obtained at $\sim$$0.1 R_B$ resolution in only $\sim$$150$\,ks, while observations of fainter targets would become feasible with either a larger effective area or longer exposures. Finally, the combination of higher angular resolution and improved sensitivity would expand the accessible sample to more than 40 LLAGNs (upper right panel in Figure~\ref{fig:bondi}), enabling the first systematic comparison of density profiles, outflow signatures, and multiphase gas structures across the LLAGN population.

Several additional capabilities would further enhance this science. Sensitivity down to $\sim$$0.1$\,keV is essential for detecting gas below Chandra's low-energy threshold, enabling constraints on the cooler phase of the hot, X-ray-emitting gas within the Bondi radius. Multiwavelength synergy is also critical: ALMA and the Event Horizon Telescope (EHT) probe the innermost accretion flow ($r \ll R_B$), JWST traces gas that has cooled out of the hot phase, and the VLA and LOFAR constrain jet power. Lynx2030 would uniquely connect these observations by measuring the hot accreting gas from Bondi-radius scales down toward the black hole, bridging the gap between the large-scale gas reservoir and the immediate vicinity of the event horizon.  Key technical requirements include an on-axis PSF of $\lesssim0.1$\arcsec\ and an effective area comparable to Chandra for density-profile measurements and hardness-ratio mapping. An effective area approximately ten times larger than Chandra would enable full temperature mapping and observations of fainter targets. A low and stable particle background is also essential for detecting the faint diffuse emission within the Bondi radius.

One important objective is to resolve the hot gas down to $\lesssim0.1R_B$ in the five best-suited targets (lower right panel in Figure~\ref{fig:bondi}). The photon requirements depend on the specific science goal. Measuring the gas density profile requires approximately 30 counts per radial bin, corresponding to exposure times of 100--350\,ks per target assuming a Chandra-class effective area. Constructing hardness-ratio maps requires about 50 counts per spatial bin, increasing the required exposure to 300--900\,ks per target with a Chandra-class effective area. Full temperature mapping is more demanding, requiring roughly 300 counts per spatial bin. For M87, this measurement could be achieved in approximately 150\,ks with an effective area ten times larger than Chandra's. Beyond detailed studies of individual systems, Lynx2030's $\lesssim0.1$\arcsec\ angular resolution would increase the number of LLAGNs with spatially resolved Bondi radii from the five systems currently accessible with Chandra to approximately 46 (upper right panel in Figure~\ref{fig:bondi}). Characterizing the hot gas within $R_B$ for this larger sample would require only $\sim$$100$--200 source counts inside the Bondi radius, with the required exposure time depending on the X-ray flux of each target.

\subsection{Gravitationally Lensed Sources}

Gravitational lensing can inform many problems: the composition of the lensing galaxy, the structure of the accretion disk around a lensed SMBH, the detection of dual/binary AGNs at high redshift, and the detection of jets and winds at high redshift. All require resolution of the individual quadruply lensed images, with typical minimum separations of $\sim 0.5^{\prime\prime}$. \textit{Chandra} has good data for about 60 quadruply lensed systems. With an order-of-magnitude improvement in sensitivity, Lynx2030 could detect $\sim 1800$ systems, compatible with the conservative numbers expected to be revealed by Rubin, SKA, and other surveys.


One of the greatest legacies of {\it Chandra} has been establishing a specific gravitational lensing configuration as one of the most powerful tools we have in all of astronomy.  The combination of strong gravitational lensing of a distant quasar by an intervening galaxy and {\bf further} lensing by the individual masses inside the galaxy (the former we call macro-lensing and the latter we call micro-lensing) allows us to perform fundamental astronomy on both the source of light (the quasar) and the lens itself (the intervening galaxy).  With this gravitational micro-lensing tool, we can measure the sizes of quasar accretion disks and coronae \citep{2007ApJ...661...19P, 2009ApJ...693..174C, 2010ApJ...709..278D, 2010ApJ...712.1129M, 2011ApJ...729...34B, 2012ApJ...756...52M, 2013ApJ...769...53M, 2014ApJ...789..125B} and probe the central engines of quasars on scales of nano-arcseconds \citep{2019BAAS...51c.487M}.  With the {\it same observations}, we can also make the crucially important measurements of the mass contents of the lensing galaxy: the stellar mass content, the dark matter content, and the stellar $M/L$ ratio at specific locations in the galaxy \citep{2004IAUS..220..103S, 2012ApJ...744..111P, 2014ApJ...793...96S}.  We can only take advantage of this tool with sub-arcsecond X-ray imaging.

{\it Chandra} has been able to very successfully perform these investigations for dozens of quadruply imaged quasars\footnote{The quadruply lensed quasars provide the best mass models for the lensing galaxies and are the focus here.}, or ``quads'' at typical redshifts of $z\sim1.5$ and lensing galaxies at typical redshifts of $z\sim0.5$.  The photon requirements are few, on the order of several $\times 10^3$ photons for image separations down to $\sim$0.4\arcsec.

Because determining stellar mass and stellar $M/L$ is \textbf{\textit{fundamental}} to much of extragalactic astronomy, getting as large a sample throughout a range of cosmic history is of paramount importance.  These same observations will also measure the sizes of quasar accretion disks and coronae for a large sample of quasars, out to redshifts of $z\sim 7$.

In addition to magnifying the flux of distant sources, gravitational lensing amplifies the spatial structure of source plane, acting as a telescope \citep{Barnacka2018} to resolve structures down to pc scales. This is the only means to probe X-ray structure on scales less than several kpc at redshifts greater than a few tenths. This technique has been used to reveal a dual AGN separated by a projected 175 pc at z=3.27 \citep{Schwartz2021}, to reveal an X-ray region 25 pc from the Gaia location of the quasar \citep{Rogers2025}, and to place ~100 pc limits on the displacement of other X-ray systems \citep{Sisk-Reynes2025}.

Simulations of the properties of the sample of quads expected to be discovered by wide-field optical surveys like Rubin predict hundreds of lensing galaxies with $z>1$ and lensed quasars with $z>5$ \cite{2010MNRAS.405.2579O, 2022AJ....163..139Y}.  As the distances increase, the separations of the quasar images tend to decrease, requiring higher angular resolution to fully exploit these systems.  As an example, recent work estimates 137 lensing galaxies with $z>1.5$ \cite{2022AJ....163..139Y}.  Only 30\% of these will have image separations $>0.5$\arcsec, and over half will have image separations $<0.3$\arcsec.

\subsection{Dual/Binary AGN and Off-nuclear AGN Phenomena}

The presence of multiple (super)massive black holes (BHs) in galactic centers naturally results from hierarchical structure formation driven by galaxy mergers \citep[e.g.,][]{1978MNRAS.183..341W}. In the typical convention \citep[e.g.,][]{2011ASL.....4..181C}, when the separation of AGN in a merger is still so large that the dynamics of the individual black holes are determined by the gravitational potentials of the galaxies, the system is a ``dual AGN''; dual pairs on Keplerian gravitationally bound orbits have been historically referred to as ``binary AGN.'' For studies of galaxy and black hole evolution, constraints on the rates of dual and binary AGN will allow us to (i) trace how supermassive black holes grow, (ii) understand the role of galaxy mergers in triggering black hole activity, and (iii) understand how galaxies assemble through major and minor mergers.

Additionally, current and near-future low-frequency gravitational wave observatories like LISA \citep{amaro-seoane2017} and pulsar timing arrays such as NANOGrav \citep{nanograv2023} will have signals strongly dependent on SMBH merger rates \citep[e.g.,][]{2022A&A...660A..68C}; as such, dual/binary AGN constraints will also (iv) predict the background signals for low-frequency gravitational wave observatories and (v) provide a better understanding of the binary AGN mergers that will be detected with LISA.

Sensitive, high-angular-resolution observations of AGN are also useful for studying offset AGN; these are a result of phenomena such as gravitational recoil (or a kick received by the resultant black hole after a merger; e.g., \cite{blecha2016}), gravitational slingshot or three-body ejection \citep[e..g,][]{uppal2024}, and off-nuclear TDEs (stellar disruptions by an off-nuclear and otherwise dormant black hole; e.g., \cite{yao2025}). Offset AGN are a critical component of a full black hole census, particularly in dwarf galaxies where as many as half of all massive black holes are off-center \citep{2019MNRAS.482.2913B}; detailed X-ray imaging will therefore also (vi) constrain the true black hole population.

Current X-ray angular resolution capabilities have been surpassed by the capabilities available in other bands of the electromagnetic spectrum. In space, HST and JWST/NIRCam offer superior angular resolution in the optical/UV and IR bands, respectively (20-200 mas range). Ground-based facilities are able to achieve orders of magnitude more powerful angular resolution than Chandra, owing to the advantages of adaptive optics and interferometry (Keck and VLTI in NIR/MIR, ALMA in sub-mm, VLBA and EHT in the radio band; 20 $\mu$as-20 mas range). Nevertheless, direct multiwavelength confirmation of sub-kiloparsec-separation dual/binary AGN via high-angular-resolution imaging remains rare \citep[e.g.,][and references therein]{fabbiano2011,trindade2024,rodriguez2006,goulding2019}. Yet even at low resolution, X-rays provide unique discovery capabilities \citep{2024ApJ...966..104S}, and the capabilities of combining matched-resolution X-ray images with multiwavelength observations \citep[e.g.,][]{2026ApJ...996...41S} will be multiplicative. 

Due to the flattening of the angular size relation, peaking at $d_A{\sim}8.5$ kpc/arcsec at $z{\sim}1.5$, half-arcsecond resolution guarantees the ability to separate pairs with few-kpc separations at \textit{all redshifts}. Further, at Cosmic Dawn ($z\gtrsim6$, $d_A\lesssim6$ kpc/arcsec), sub-kpc separations become resolvable as the resolution moves below $0.2$\arcsec. Given the small sizes observed for such high-redshift galaxies \citep[e.g.,][]{2024ApJ...960..104S, 2026ApJ...999L...6M}, the increase in resolving capabilities from 0.5\arcsec\ to 0.1\arcsec\ represents a fundamental change from exploring dual AGN in separate galaxies to those that are now within the same galaxy. Likewise, in the local Universe, a nominal 0.1\arcsec angular resolution would allow us to resolve dual and binary AGN at separations as small as $\lesssim 20$ pc at $z \lesssim 0.01$.

\textit{Chandra} results have paved the way for fully exploiting available angular resolution. For CCDs, Energy Dependent Subpixel Event Repositioning \citep[EDSER;][]{2003ApJ...590..586L} and Bayesian modeling of point source imaging \citep{2019ApJ...877...17F} push discovery space beyond nominal resolution limits. Multiwavelength synergies with telescopes like HST and JWST are capable of identifying stellar nuclei, merger structures, dust lanes, and host centroids \citep[e.g.,][]{Li2024}, while long-baseline interferometry is capable of pinpointing impostor radio jet structures \citep[e.g.,][]{walsh2023}. For Lynx2030, hard X-ray response ($2-10$ keV) is important for detecting AGN continuum and iron line emission, as well as for avoiding intrinsic absorption; likewise, soft response $\lesssim 2$ keV is important for detecting off-nuclear TDEs.

With \textit{Chandra}, meaningful constraints on tightly-separated binaries requires more than ${\sim}20$ counts for the fainter source \citep{2020ApJ...892...29F}, while fewer than 100 total combined counts fail to meaningfully constrain pair parameter space \citep{2022MNRAS.514.2855P}. That said, even as few as 3 counts can be meaningful for slightly larger separations \citep{2019ApJ...887..171C}; the key requirement is that both sources produce enough counts to be meaningfully separable in probability space.

\subsection{AGN Jets}
\begin{figure}[h]
\centering
\includegraphics[width=0.54\textwidth]{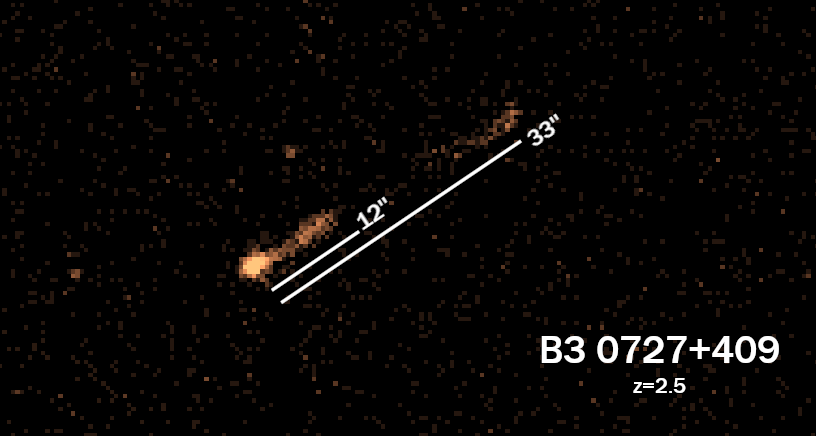}
\includegraphics[width=0.45\textwidth]{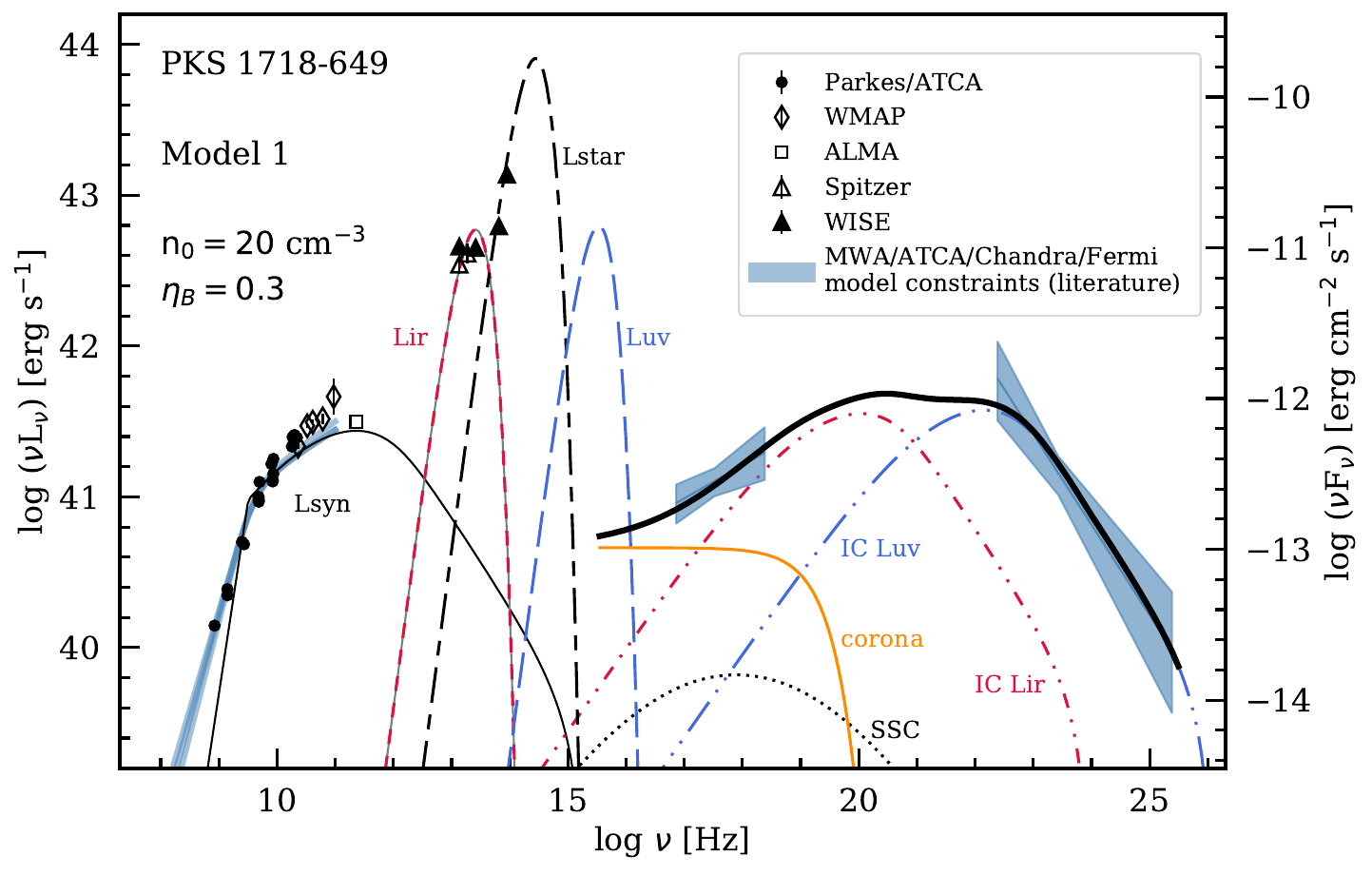}
\caption[X-ray constraints on jet emission mechanisms]{\textbf{Left:} \textit{Chandra} view of $z=2.5$ AGN B3 0727+409 and its jet (\cite{2023ChNew..33....1S}; adapted from \cite{2016ApJ...816L..15S}). Extending over 100 kpc in projected length ($d_A=8.225$ kpc/arcsec at $z=2.5$), this serendipitously detected source has no radio counterpart, strongly suggesting an IC/CMB interpretation. \textbf{Right:} Emission models for the compact radio source PKS 1718-649, which is both young ($\sim 100$ yr) and close ($z = 0.014$), enabling a detailed study of the initial radio jet expansion \citep{sobolewska2022}. While gamma-ray constraints are well matched by models, X-ray emission is contaminated by structure of size 10--100 pc, motivating higher-angular-resolution observations.
}
\label{fig:high-z_jet}
\end{figure}  

Beginning with an accidental discovery in focusing observations with \textit{Chandra} \citep{2023ChNew..33....1S}, detailed imaging observations of X-ray emission from relativistic jets powered by AGN have presented challenges for interpretation \citep{2000ApJ...542..655C, 2000ApJ...544L..23T}. Radio observations have firmly established that jets are populated by relativistic particles \citep[e.g.,][]{1980ARA&A..18..165M}, with radio emission arising from synchrotron interactions of these electrons with magnetic fields. However, the emission mechanisms responsible for the high-energy emission of AGN jets are still unclear, with conflicting lines of evidence supporting independent and source-specific interpretations (e.g., conflicting results from both \cite{2017ApJ...835L..35M} and \cite{2019ApJ...883L...2M}).

X-ray emission mechanisms typically fall into two categories: those relying on a single population of jetted particles and those invoking a second population, whether leptonic (electrons) or hadronic (protons; e.g., \cite{2002MNRAS.332..215A}), to produce models capable of satisfying observed radio, optical, X-ray, gamma-ray, and polarization constraints \citep[e.g.,][]{2020ApJ...893...41W}. For a single population, synchrotron self-Compton (SSC) of the radio synchrotron emission to X-ray energies has long been held infeasible to explain observed X-ray fluxes \citep{2000ApJ...540L..69S}; rather, the Cosmic Microwave Background (CMB) is invoked to supply the necessary seed photons for inverse Compton scattering to X-ray energies \citep[IC/CMB;][]{2000ApJ...544L..23T}.

For investigating the contributions of IC/CMB emission, there is much potential at $z>2$, particularly at $z>5$. Owing to the availability of photons from the CMB---the energy density of which scales as $U_{\rm CMB}\propto (1+z)^4$---the relative strength of X-ray to radio is expected to increase, as IC/CMB becomes a dominant cooling mechanism over synchrotron. This interpretation is bolstered by the detection of X-ray jets with no radio counterparts (Figure~\ref{fig:high-z_jet}; \cite{2016ApJ...816L..15S}, \cite{2021ApJ...911..120C}), the evolution of $L_X/L_R$ ratios with redshift \citep{2011ApJ...726...20M, 2021MNRAS.505.4120I}, and direct SED constraints, when available \citep{2022A&A...659A..93I}. The latter are best achieved with blazars, where fluxes are boosted to improve signal-to-noise; in these cases, excellent angular resolution is needed to clearly separate the jet from the core. For non-blazars, typical radio jet lengths at high redshift extend no more than ${\sim}1$ kpc \citep{2018ApJ...861...86M}; here, again, high angular resolution will permit direct radio--to--X-ray SED constraints. Yet, even without radio detections, jets are potentially key ingredients in growing the first supermassive black holes \citep{2024Univ...10..227C}, and the ability to search for extended X-ray structures in the innermost kpc around high-redshift quasars, both radio-loud and radio-quiet, will be key to investigating if jets are common enough to meaningfully impact earliest black hole growth.

At low redshift, there is significant potential in investigating compact double radio sources with sub-kiloparsec projected linear sizes, which are believed to represent the early stages (the first few hundred to a few thousand years) of an AGN jet expansion through its host's interstellar medium. External Compton scattering occurring in the compact radio lobes has been proposed as the origin of their X-ray emission (e.g., \cite{stawarz2008}, \cite{ostorero2010}, \cite{sobolewska2022}, \cite{krol2024}; shown in Figure \ref{fig:high-z_jet}). Current X-ray angular resolution capabilities do not allow us to resolve emission corresponding to the lobes from that of the core, confirm that the compact radio lobes emit X-rays, and uncover the conditions governing the earliest stages of AGN-galaxy feedback related to jet launching and initial jet expansion. However, a $\sim$ 0.1\arcsec\ X-ray angular resolution would allow us to resolve lobe emission from core emission in 12 (of 79) sub-kpc scale jets in the local Universe \citep[angular sizes $\sim$100--280 mas; see the sample of][]{kiehlmann2024} and directly address the question of whether the compact radio lobes emit X-rays via the external Compton mechanism. High-angular-resolution radio surveys continue to discover new compact and young AGN jets in the nearby Universe \citep[e.g., 5 more suitable 109--157 mas-scale targets in the sample of][]{sheldahl2025}, so this sample may be much larger by the launch of Lynx2030.

For this science case, synergies with radio are critical, as well as, to a lesser extent, sub-mm, FIR, NIR, and optical resolved jet images and higher-energy flux measurements. Particularly at radio frequencies, jets have been resolved into complex structures, including lobes, knots, and hot spots, for decades \citep[e.g.,][]{1974MNRAS.167P..31F}; for X-rays, multiple resolution elements per jet are required to begin matching structures across wavelength domains, while having multiple resolution elements per jet component will enable direct tests of the one- or two-population model question \citep[e.g.][]{2023ApJS..265....8R}. For the unique case of M87, \citet{2026arXiv260613800P} recently used deconvolution and long temporal baselines to demonstrate some of the potential that few-tenths-of-an-arcsecond imaging will enable for multi-wavelength jet studies.

\subsection{IMBHs vs. ULXs in Dwarf and Metal-Poor Galaxies (High-z Analogs)}

In driving the investigations of X-ray sources that are relevant to the early Universe, nearby low-metallicity dwarf galaxies have served as important analogs, despite their relative rarity in the X-ray accessible Universe \citep[$z<0.1$; see, e.g.,][]{Morales-Luis2011}. Due to less mass and angular momentum loss through stellar winds at low metallicities, X-ray binaries produced in these environments are expected to be more numerous and luminous \citep[e.g., ultraluminous X-ray sources, ULXs, with $L_{\rm X} \gtrsim 10^{39}$ erg s$^{-1}$;][]{Mapelli2010, Prestwich2013, BZ2016}. This leads to an inverse relationship between the galaxy-integrated luminosity per star formation rate (SFR) and the metallicity, referred to as the $L_{\rm X}$-SFR-$Z$ relation \citep[see][and references therein]{BZ2013,  Brorby2016, Lehmer2022}. 

Recently, \citet{HERA2023} observed the imprint left by X-ray sources in the early Universe via the 21-cm signal from the Epoch of Reionization, finding that the $L_{\rm X}$-SFR-$Z$ relation based on nearby low-metallicity galaxies agrees well with constraints placed on the X-ray power from $z>6$ galaxies. However, at the lowest metallicities ($Z<$10\% $Z_{\odot}$), corresponding to the average metallicity at these early cosmological epochs, the $L_{\rm X}$-SFR-$Z$ relation is poorly constrained for a number of reasons, including scatter due to stochastic variation for sources within the galaxies, and poor statistics due to few extreme metal-poor galaxies existing within the X-ray accessible volume.  The solution in both cases is to {\it increase adequate sampling of these galaxies}, which is only possible with more sensitive X-ray telescopes, capable of surveying a statistically significant population of dwarf galaxies. 

Complicating matters further: dwarf ($M_\star < 10^{9} M_{\odot}$) galaxies, which dominate the lowest metallicity population, are very compact (extending about a few arcseconds) or exhibit irregular morphologies. Accreting intermediate-mass black holes (IMBHs) in these galaxies would have similar X-ray luminosities \citep{Wang2025, Cann2020, Cann2024}, may still be settling and therefore not located at the nuclei of the galaxies \citep{Reines_2020_radio, Thibaut2024}, and/or may be co-located with X-ray binary sources \citep{Matzko2026, HK2020}. Therefore, having sub-arcsec ($\lesssim$ 0.5\arcsec) angular resolution is paramount for identifying potential IMBHs and disentangling their contribution from X-ray binary emission. 

To measure the contributions from IMBH and XRB sources, multiwavelength spectro-photometric data (spanning X-ray to infrared) that have $\lesssim$0.5\arcsec\ resolution are necessary. While the multiwavelength coverage capability is currently available (and growing with upcoming observatories), correspondingly sensitive X-ray data are not, without impossibly long exposure times. This work requires an X-ray telescope that can obtain $\sim$ 100 counts for 0.5 -- 8 keV fluxes $\sim 10^{-16}$ erg s$^{-1}$ cm$^{-2}$ within reasonable exposure times ($\sim 10 - 30$ ks) in order to survey the XMPG population ($\sim$ hundreds at $z<0.3$) in a statistically meaningful way, and further push X-ray observations to greater cosmological volumes ($z\sim1$).

\subsection{SMBH and XRB in M31 \& Other Local Galaxies}

Despite our location in a relatively rural area of the universe, the Local Group contains a rich collection of galaxies hosting diverse populations of X-ray binaries (XRBs) and known or candidate supermassive black holes (SMBHs). An X-ray telescope with $<0.5$\arcsec\ spatial resolution will expand a long campaign of monitoring X-ray variability in nearby galaxies while disentangling sources in crowded central regions.

As the nearest large galaxy outside our own and the only other large galaxy in the Local Group, the Andromeda galaxy (M31) is a unique laboratory to investigate the evolution and composition of field spiral galaxies up-close. Like the Milky Way, it contains a substantial population of X-ray binaries and a central SMBH \citep{2010ApJ...710..755G,2002ApJ...577..738K}. Our high-inclination view reveals the entire spiral disk and bulge of M31, unlike our obstructed view of the Milky Way's structure, allowing a top-down study of M31's central region and a direct view of the SMBH at its center. Like Sgr A*, M31* is a very low-luminosity SMBH, only visible because of its close proximity.

The central region of M31 contains a supermassive black hole (M31*) and several X-ray binaries all within a central $\lesssim 1$\arcsec\ ($\approx 4 \:\rm{pc}$ at a distance of $750 \:\rm{kpc}$). Currently, only \textit{Chandra} can investigate the X-ray emission from this crowded region, but only via 2D image reconstruction \citep{2011ApJ...728L..10L,2025ApJ...981...50D}. Extracting a clean X-ray spectrum of M31* is normally impossible with 1\arcsec\ spatial resolution; improved $<0.5$\arcsec\ resolution will cleanly disentangle each nuclear X-ray source. A telescope that exceeds \textit{Chandra's} capabilities will constrain activity and flares from M31* and enable direct comparisons with Sgr A* and other low-luminosity SMBH systems.

Outside of its core region, M31 also hosts a rich population of X-ray binaries \citep{1993ApJ...410..615P,2002ApJ...577..738K,2011A&A...534A..55S} previously observed with a range of X-ray telescopes. Increased spatial resolution will more accurately disentangle confused point sources and coincident diffuse emission and allow for more complete comparisons with the X-ray binary population in the Milky Way.

Besides the two large spirals in the Local Group, other smaller local galaxies also host populations of X-ray binaries and may also have SMBHs or intermediate-mass black holes at their centers. M32, the only elliptical galaxy in the Local Group, has a central cluster of XRB and a candidate SMBH within a few arcseconds of the optical centroid of that galaxy \citep{2003ApJ...589..783H}, and the Triangulum galaxy M33 \citep{2005ApJS..161..271G} also has a substantial population of X-ray binaries. Because of the proximity of these and other satellite and dwarf galaxies in the Local Group, their populations of X-ray sources can be studied with much greater precision than in the more distant universe.

By improving on \textit{Chandra}'s currently state-of-the-art $\lesssim 1$\arcsec\ angular resolution, a future X-ray telescope with $<0.5$\arcsec\ resolution will primarily benefit science cases that struggle with source confusion, overlap, or crowding. At the edge of the Local Group at $d \approx 1 \:\rm{Mpc}$, this spatial resolution can discern point and extended sources at separations of $r \approx 2 \:\rm{pc}$. 

In M31, for example, this resolution will allow extraction and modeling of X-ray spectra from M31* and the candidate SMBH in M32 by disentangling the confused and coincident X-ray sources in the crowded central region of M31's bulge. In other Local Group galaxies such as the spiral M33 and the elliptical M32, this capability will enable detailed studies of extended X-ray emission regions and X-ray binary point sources, as well as the candidate SMBHs tentatively identified in the dense central regions of several Local Group galaxies.

A large field of view ($\sim 0.3^\circ$) would facilitate surveys of every small Local Group galaxy in a single pointing, and of M31 in at most six pointings. This is crucial for efficient time-domain surveys of variability of supermassive black holes and X-ray binaries in local galaxies that appear too large on the sky compared to more distant targets.

Many of the X-ray point sources in Local Group galaxies emit most of their high-energy emission via continuum processes close to $\sim 1 \:\rm{keV}$, but other targets like dense molecular clouds reflecting external photon or cosmic ray irradiation require substantial energy response at $6.4 \:\rm{keV}$. Therefore, an energy range spanning $0.1 - 10\:\rm{keV}$ is a valuable complement to increased spatial resolution.

\subsection{Pulsars and Binary Stellar Systems in Globular Clusters}

Because of their high rates of stellar encounters, globular clusters (GCs) are the most efficient ``factories'' in the Universe for the production of exotic, close binary systems comprising either a normal star and a compact object or two compact objects.  GCs have often been cited as potential breeding grounds for the progenitors of double-degenerate Type Ia supernovae and of neutron star-neutron star (NS-NS) systems that will eventually merge \citep[e.g.][]{2006NatPh...2..116G, 2009A&A...498..329G, 2010ApJ...720..953L, 2013ApJ...776...18F, 2014ARA&A..52...43B}; such systems featured heavily in the Astro2020 Decadal Survey and have been the subject of hundreds of papers.

A better understanding of the internal dynamics of GCs and the dynamical formation of exotic binary systems is a longstanding and crucial problem in astronomy;  however, the details of GC close binary formation are largely uncertain. Theoretical work on GC-produced exotica relies on high-throughput numerical simulations of GCs; the fidelity of these simulations is best tested observationally with direct tracers of internal dynamics.  Low-luminosity X-ray sources in Galactic GCs, discovered in great numbers with {\it Chandra}, are a heterogeneous mix of close binaries and their progeny, and are {\bf the best tracers of GCs' internal dynamics}.

Early, deep  {\it Chandra} observations of
about a dozen GCs not only discovered hundreds of low-luminosity X-ray sources but also allowed for initial progress in classifying the members of this low-$L_X$ population as either quiescent LMXBs (qLMXBs), cataclysmic variables (CVs), millisecond pulsars (MSPs), or chromospherically active main-sequence binaries (ABs).  All are close binary systems or, in the case of MSPs, the progeny of close binary systems.  Thus, it was recognized early on that sensitive, high-resolution X-ray observations were extremely efficient at finding the ``needles in a haystack'' close binaries.

Each sub-population probes the GC internal dynamics in a different way.  For example, the qLMXB-dominated population has a very strong dependence on encounter frequency \citep{2003ApJ...591L.131P, 2003ApJ...598..501H, 2003A&A...400..521G} while a CV-dominated population has a weaker dependence \citep{2006ApJ...646L.143P}, suggesting the CV population had both primordial and dynamical contributions.  It was suggested that the ABs in M4 are a largely primordial population \citep{2023MNRAS.524.2088L}, but this was based mainly on a comparison with only one other well-studied cluster, NGC 6397, as these are two of the closest GCs and some of the only ones (along with 47 Tuc and NGC 6752) whose close binary populations have been probed well below $10^{30}$ erg/s. 

A thorough understanding of GC internal dynamics can be obtained with complete samples of the various close binary populations in a large sample of GCs.  These sources are both faint and crowded in GC cores.  The faintest sources are the ABs, whose luminosity function goes down to $\sim10^{28}$ erg/s \citep{1997ApJ...478..358D}.  Some of the most crowded sources can be seen in the collapsed core of NGC 6397 ($d = 2.3$ kpc), with separations as small as $\sim$ 0.5--1\arcsec.  A representative sample of GCs will require observations out to typical GC distances of $\sim$8~kpc \citep{2021MNRAS.505.5957B}, at which the fluxes of the faintest sources would be $\sim 10^{-18}$ erg/cm$^2$/s, and the source separations would be $\sim$ 0.15\arcsec.

\subsection{Stars}

Young Stars, Zero-age Main sequence (ZAMS) and prior, produce copious X-rays.  The production mechanisms vary.  Hot stars (OB) produce X-rays via high-velocity winds that interact with the ISM, or via collisions with winds from nearby stars; in these cases, L$_x$/L$_{bol}$ is about 10$^{-7}$.  Young low-mass stars (G-M) generate X-rays primarily through the breaking of magnetic field lines.  In these cases, L$_x$/L$_{bol}$ is about 10$^{-3}$. X-rays from cool stars probe the critical interface between the star and its disk and envelope while they are present \citep{Shu1997}.  However, the strong magnetic fields and X-rays persist for tens to hundreds of millions of years longer than the disk.  This allows us to use X-rays as a marker of youth and as a mechanism to study accretion and the stellar interior.  

Multiple studies can be carried out in star-forming regions. The first is a cluster census.  This allows us to break the populations down into the fraction of disked versus undisked stars - a study impossible without X-rays. From this, we can derive disk survival times as a function of spectral type (with deep enough X-ray studies). X-rays are seen to impact cluster morphology and destroy grains.  More recent work is focused on the effects of X-rays on planet formation, especially the impact of X-ray flares on young planets \citep{2017ApJ...847...29O}.  

Work to date is highly incomplete. Few exoplanet hosts are well characterized, especially young exoplanet hosts.  Fewer than 5 Class 0 protostars have been detected \citep{Grosso2020}; these and similar detections will give insights into how X-rays guide the earliest stages of star formation. 

Only one star-forming region has had a complete survey: the core of the Orion Nebular Cluster \citep{Getman2005}.  This was a very productive observation. But Orion is a single intermediate-density cluster; the stars in the survey are all within 2 million years of the same age and have the same initial conditions. The data strongly hint at evolutionary characteristics affected by the star's natal source, including disk survival time, multiplicity, and the seeding of disks with modified material from local supernovae.

For intermediate and massive clusters further than 400 pc from the Sun, 0.5\arcsec\ resolution is insufficient to resolve each stellar system.  Even at the distance of Orion, most multiple systems are not resolved at a level that is routine from ground-based instruments (e.g., 0.1\arcsec\ from ALMA and adaptive optics).

There are 36 embedded clusters located within a kpc of the Sun \citep{Gutermuth2009}. There are an additional 17 massive clusters within 3.5 kpc of the Sun \citep{Kuhn2014}. 

\begin{figure}
    \centering
    \includegraphics[width=0.5\linewidth]{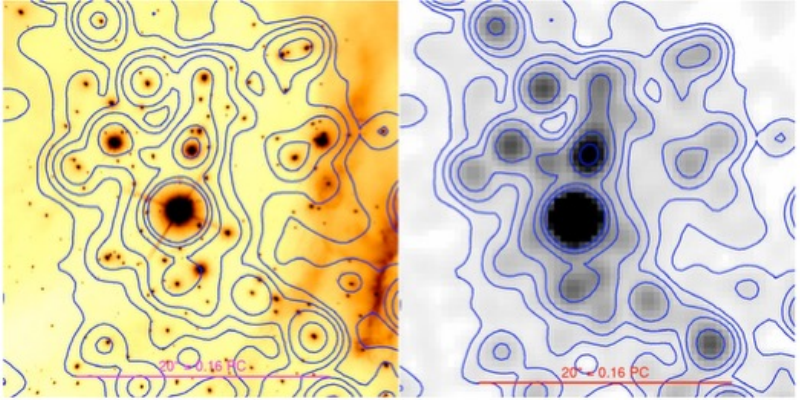}
    \caption{\textbf{Right:} A Chandra image of RCW 38 at 1.7 kpc with 0.5\arcsec\ resolution; contours come from the X-ray data.  \textbf{Left:} an image of RCW 38 at 0.1\arcsec\ resolution. Several of the X-ray sources are now understood to be composed of multiple stars, and some, but not all, of the diffuse emission resolves into stars.}
    \label{fig:RCW 38}
\end{figure}

The study of Star formation is fundamentally multiwavelength. In 2000, Chandra matched the resolution of the best available ground- and space-based telescopes.  JWST, ALMA, and the E-ELT are all pushing to higher resolution.  This means that physical sources of the X-rays remain poorly understood as we push out toward the center of the galaxy, as well as star-forming regions in nearby galaxies (e.g., 30 Doradus in the LMC).  

\begin{figure}
    \centering
    \includegraphics[width=0.5\linewidth]{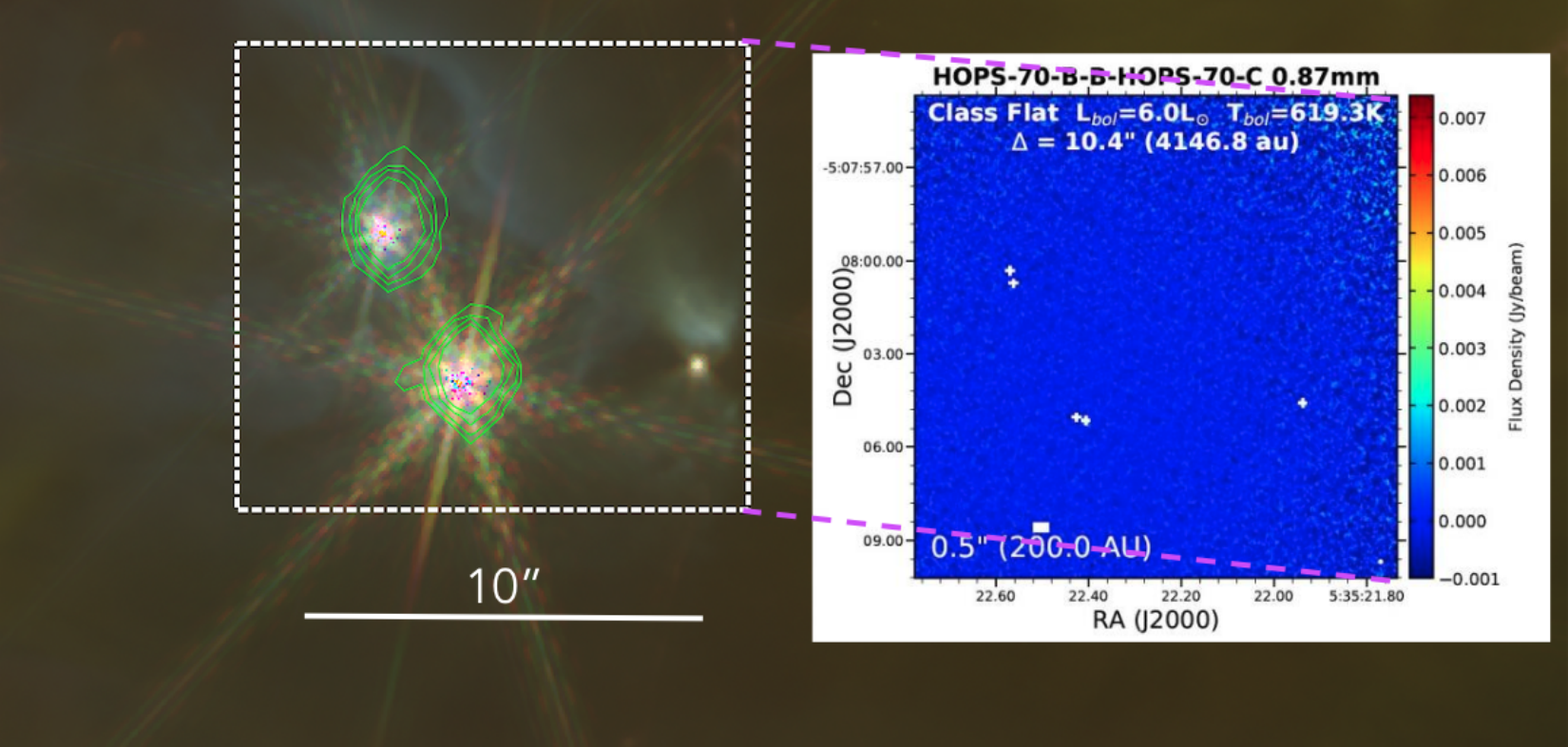}
    \caption{\textbf{Left:} A JWST NIRCam image of HOPS-70.  The white box is about 14$^{\prime\prime}$ on a side. The green contours indicate Chandra flux from a 100~ks observation. \textbf{Right:} An ALMA 0.87 mm image of the boxed region. The two X-ray sources are further resolved into binaries separated by $\sim 0.25^{\prime\prime} $ (100 AU). https://planetstarformation.iaa.es/HOPS-70}
    \label{fig:HOPS-70}
\end{figure}

   About 100 photons per star is sufficient to measure the gas along the line of sight and perform a one-temperature fit to estimate the X-ray luminosity. The lowest-mass young stars have X-ray luminosities of about 10$^{27}$ erg/sec.

\subsection{Enabling Techniques and Technologies}

\subsubsection{Adaptive Spectral Optimization for Enhanced Source Separation}

Crowded X-ray environments remain challenging even at sub-arcsecond angular
resolution because overlapping point-spread functions (PSFs), low photon
counts, background emission, and high source densities can produce ambiguous
source assignments. Although improved angular resolution reduces source
confusion, spatial information need not be used in isolation. Individual
X-ray events also carry energy information, and sources with distinct spectral
energy distributions may remain spectrally distinguishable even when their
spatial profiles partially overlap. Methods that jointly exploit spatial and
spectral information have demonstrated the value of energy-resolved information
for disentangling overlapping X-ray sources and probabilistically assigning
detected events to their most likely source of origin
\citep{Broos:2010,Meyer:2021}.

Building on established use of energy-resolved information, we propose
investigating whether a compact set of energy bands can be optimized to
complement angular resolution in separating neighboring or partially overlapping sources. Rather than relying exclusively on predetermined broad energy intervals, candidate band configurations could be evaluated according to the expected spectral contrast between relevant source populations, instrumental response, background conditions, PSF overlap, and counting statistics. Band placement and width could be considered jointly, since narrower bands may isolate diagnostically useful spectral structure while broader bands provide improved photon statistics. The resulting measurements would provide a low-dimensional spectral representation designed specifically for source-separation performance.

An optimized configuration need not consist exclusively of bands that are individually maximally discriminative. Different energy intervals may instead provide complementary information. Bands containing strong spectral contrast between candidate source populations could provide source-sensitive measurements, while other intervals could characterize continuum emission or spectral components shared between sources. Optimization should therefore consider the information carried jointly by the complete band configuration, rather than ranking individual energy intervals independently.

Energy-band measurements and hardness ratios are established tools for
characterizing X-ray spectral properties \citep{Evans:2024}. Building on these approaches, normalized differences, ratios, or other combinations of optimized bands could be investigated as features for crowded-source discrimination.
Such derived features may reduce sensitivity to overall source brightness while preserving differences in spectral shape. They should be evaluated as complementary to, rather than replacements for, the original band measurements, since absolute and relative measurements may contain different information about the underlying sources.

The optimization procedure should be separated from its validation. A
candidate spectral configuration could first be optimized using a designated simulation or training population and then fixed before evaluation against independent source populations, instrumental-response realizations, background levels, PSF configurations, and photon-count regimes. Independent validation of a fixed spectral configuration would help determine whether improved source separation reflects generalizable spectral information rather than optimization specific to a particular simulated population. Comparisons should
include conventional fixed-band strategies and, where computationally
feasible, methods that retain the full available event-energy information.

Robustness to nuisance variables should also be evaluated explicitly.
Variations in background spectrum, source spectral parameters, intervening absorption, detector response, calibration, and PSF structure may alter the apparent separability of candidate sources. Simulations that vary these quantities individually and jointly could identify degeneracies between source properties and observational conditions and determine which spectral measurements retain discriminatory information after these effects are introduced. This would also help establish whether a band configuration
optimized under one set of conditions remains useful across a broader range of realistic observing scenarios.

Potential applications include Galactic centers, dual and binary active galactic nuclei (AGN), dense stellar clusters, and deep extragalactic surveys. Adaptive spectral optimization would complement, rather than replace, improvements in angular resolution. Increasing angular resolution remains the primary means of reducing spatial source overlap and is critical for avoiding source confusion in deep X-ray observations \citep{Gaskin:2019}. Optimized spectral information could provide an additional dimension of discrimination when source PSFs nevertheless remain partially overlapping, helping to assign photons or flux to individual sources. This combination may be particularly valuable in photon-limited observations, where source detection and characterization are strongly affected by Poisson statistics and uncertainty in photon assignment \citep{Broos:2010,Meyer:2021}.

Future investigations should incorporate realistic astrophysical source populations, energy-dependent PSFs, detector response matrices, effective-area curves, and astrophysical and instrumental backgrounds. Studies spanning low-count (10--100 counts per source) and moderate-count ($>100$ counts per source) regimes could quantify the trade between spectral discrimination and photon statistics. Evaluate performance using metrics including source-recovery fraction, localization error, photon-assignment accuracy, classification performance, contamination between neighboring sources, and changes in the practical confusion limit.

Treat uncertainty as an explicit component of the analysis rather than reporting only point estimates or classification labels. Posterior source-assignment probabilities and other calibrated measures of uncertainty can help identify regimes in which source separation becomes ambiguous as photon counts decrease or source overlap increases \citep{Meyer:2021}. Uncertainty-aware evaluation would provide a more realistic measure of the scientific benefit of adaptive spectral optimization and help identify the observing regimes in which spectral information most complements Lynx's angular resolution.

\begin{table}[p]
\centering
\caption{Science Case and Enabling Techniques Summary (including non-pillar cases, not discussed in the text).}
\label{tab:science_summary}
\scriptsize
\renewcommand{\arraystretch}{1.15}
\setlength{\tabcolsep}{4pt}
\begin{tabular}{|L{1.9cm}|L{3.5cm}|L{3.6cm}|L{2.6cm}|L{3.3cm}|}
\hline
\rowcolor[HTML]{203864}
\color{white}\textbf{Case} &
\color{white}\textbf{Question} &
\color{white}\textbf{Measurement} &
\color{white}\textbf{Observable} &
\color{white}\textbf{Photons/Effective Area} \\
\hline
AGN feedback I &
How does AGN energy couple to diffuse gas in massive systems, and how has this evolved since cosmic noon? &
Image shocks, cavities, cold fronts, and outflows out to $z\sim2$ &
0.5--7 keV morphology &
$>$20,000 counts within 10~kpc; $>$10 counts/pix for small cavities \\
\hline
AGN feedback II &
How does AGN power couple to the ISM in non-radio-loud AGNs? &
Resolve nuclear/jet regions; spatially resolved spectra &
0.3--7 keV morphology; soft-X-ray lines; hard continuum/Fe~K &
$\sim100\times$ \textit{Chandra} area; $\sim$46,000 counts (inner $1.5''$), $\sim$23,000 (cone); $\gtrsim$10 counts/beam (morphology), 50--100 (spectra) \\
\hline
BH seeds &
What seeds the first BHs? &
Detect and localize faint accreting BHs at $z\gtrsim8$; high-$z$ AGN demographics &
0.3--10 keV flux, hardness, positions &
10--20 counts for detection/localization; 50--100 for basic spectra \\
\hline
Resolving the Bondi radius &
How are AGNs fueled? Is the accretion flow anisotropic/multiphase, with outflows? &
Image Bondi regions of nearby SMBHs; density, hardness-ratio, and temperature maps &
0.1--2 keV flux and spectra &
30 counts/radial bin (density); 50/bin (HR maps); 300/bin (temperature); 1--10$\times$ \textit{Chandra} area \\
\hline
Dual/binary and off-nuclear AGN &
How do BH pairs, recoils, and off-nuclear accretors trace hierarchical growth? &
Resolve sub-kpc and, locally, tens-of-pc separations &
Point-source positions, spectra, variability &
Source-dependent; confusion limited \\
\hline
SMBH and XRB in M31 and local galaxies &
How do SMBHs form and evolve in Local Group galaxies? How do XRB populations compare? &
Spectroscopy and imaging of crowded galactic centers out to $\sim$1~Mpc &
Flux, spectra, variability &
100 counts in $\sim$10~ks \\
\hline
IMBH vs.\ ULX in high-$z$ analogs &
What powers X-ray emission in rare, local early-Universe analogs? &
Sub-arcsec imaging and X-ray SEDs of point sources in galaxies &
0.5--7 keV flux &
100 counts at $\sim10^{-16}$ erg s$^{-1}$ cm$^{-2}$ \\
\hline
AGN jets &
What produces jet X-rays across redshift? Where are jets launched? How is energy transported? &
Resolve jets at radio scales; disentangle point-source emission; spectral features; multi-year time evolution &
Morphology, spectral indices, flux, time evolution &
Source-dependent; $\sim$100 counts/feature for spectral modeling \\
\hline
Pulsars and X-ray binaries in star clusters &
How do dense star clusters form exotic close binaries, and what is their role in Ia SNe and NS--NS merger progenitors? &
X-ray source populations in dozens of globular clusters &
Sources detected above $10^{28}$ erg s$^{-1}$ &
$\sim$5--10 photons at $\sim10^{-18}$ erg s$^{-1}$ cm$^{-2}$ \\
\hline
Stars &
What is the density of star clusters? How long do disks survive? How do flares impact exoplanet atmospheres? &
Resolve all stars in open clusters out to the LMC/SMC ($0.1''$) &
0.5--7 keV flux &
Completeness at LMC/SMC: $\sim10^{-19}$ erg s$^{-1}$ cm$^{-2}$ \\
\hline
Lensed sources &
What are the stellar mass content and M/L of lens galaxies, and the sizes of quasar emission regions and coronae? Do they evolve over cosmic time? &
Quadruply lensed quasars (lens galaxies at $z\approx0.5$--3; quasars at $z\sim1$--7) &
X-ray fluxes of individual quasar images &
$\sim10^3$--$10^4$ photons \\
\hline
Adaptive spectral optimization &
Can optimized spectral information complement angular resolution for separating partially overlapping X-ray sources? &
Improved source localization, reduced confusion, and improved classification or photon assignment for overlapping sources &
Energy-resolved measurements in a compact set of optimized, complementary bands, including normalized spectral features &
Simulation-based; performance depends on spectral contrast, photon statistics, source density, PSF overlap, background, and instrumental response; requires independent validation of fixed band configurations \\
\hline
\end{tabular}
\end{table}

\section{Energy Bandpass Working Group}
\sectionrule

The Lynx2030 SAG Bandpass Working Group is charged with defining the best energy range for a next-generation flagship X-ray observatory in order to meet its science objectives. These include identifying and characterizing the first black holes, elucidating the co-evolution of AGN and their host galaxies, probing the formation of large-scale structure in the Universe, and revealing the physics behind the high-energy transient phenomena. To achieve these goals, such a mission requires a PSF comparable to Chandra, coupled with $\sim 100 \times$ its effective area and a large FoV. We recommend a notional bandpass of $\sim 0.1-12$ keV. The upper bound ensures that the Fe K band and Compton hump are captured for AGN at $z > 6$ while also maintaining enough energy coverage to perform detailed observations of nearby sources. The lower bound of this energy range facilitates detailed studies of AGN outflows at cosmic noon and enables the investigation of low-temperature diffuse emission from early galaxy clusters and groups.

\subsection{Primary Science Drivers}
 
Our recommendation is motivated by four core scientific pillars that drive bandpass requirements for a next-generation X-ray flagship mission:
 
\subsubsection{Discovery of Early Black Holes}
 
By combining high angular resolution with a massive increase in effective area, the mission will unambiguously detect the seeds of supermassive black holes at $z \sim 6-10$, providing a needed synergy with IR observations from JWST and Roman as well as a crucial benchmark for numerical simulations. Using the Fe K band and Compton Hump as critical observables, this science requires the effective area to reach its peak between $\sim$0.6-3 keV. Extending the low-energy response down to $\sim$0.1 keV will capture redshifted soft X-ray emission from rapidly accreting, low-mass SMBH seeds, improving sensitivity at the faint end of the high-redshift X-ray luminosity function and constraining obscuration and outflowing winds in AGN at cosmic noon.
 
\subsubsection{AGN-Host Co-evolution}
 
Utilizing high-resolution X-ray spectroscopy via a microcalorimeter and/or gratings, the mission will probe the physical mechanisms behind how galaxies and their central black holes grow together across cosmic time through accretion and outflows. This will expand on the legacies of XRISM and the gratings on Chandra and XMM-Newton, and will complement multi-wavelength observations from GMT, E-ELT, JWST and ALMA. Using emission and absorption features from O VII, O VIII, and C V, as well as the Fe K band and Compton hump as proxies, this science requires a bandpass of $\sim$0.1-12 keV to capture all of these features in AGN out to z $\sim$ 3. Sensitivity down to $\sim$0.1 keV is essential for retaining redshifted soft-X-ray diagnostics of warm absorbers, photoionized outflows, and the soft excess, thereby connecting SMBH accretion and feedback to host galaxy evolution.
 
\subsubsection{Formation of Large-Scale Structure}
 
By observing the high-redshift universe, the mission will map the growth of the earliest galaxy clusters and groups, as well as the WHIM between and around galaxies, tracing the missing baryons that bridge this expanse. High-resolution imaging will map the gas density, where asymmetries and filamentary structures reveal the violent assembly of early protoclusters, in contrast to smoother, virialized systems. High-resolution spectroscopy will measure the gas temperature, a direct proxy for the gravitational potential. Characterizing these thermal profiles at $z > 5$ is essential for benchmarking cosmological simulations. Emission lines from Fe XXV and XXVI, O VII and VIII, and Si XIII and XIV will reveal the presence of heavy elements forged in early supernovae, providing a window into the star-formation history and the enrichment of the ICM over cosmic time. This science requires sensitivity down to 0.1 keV to perform detailed line diagnostics on the above ions out to z $\sim$ 5 and beyond.
 
\subsubsection{TDAMM Science}
 
High-sensitivity observations of \textbf{LISA X-ray counterparts} from pre- and post-merger accretion disks from z $\sim$ 1-4 will illuminate the connection between SMBHs and their host galaxies, revealing binary SMBH populations and early seed evolution. Rapid follow-up of \textbf{neutron star mergers detected by LIGO} will shed new light on GRB afterglows, probing jet launching and compact remnant properties. High-sensitivity soft X-ray coverage ($<$0.2 keV) will also probe mass transfer dynamics, accretion energy dissipation, and magnetic field strengths in \textbf{interacting white dwarf binaries}, clarifying their potential as Type Ia SN progenitors. Soft X-ray observations (0.1 keV) of \textbf{TDEs} detected by facilities such as Rubin and UVEX will probe accretion physics, disk formation, and super-Eddington winds via high-resolution measurements of ions from C, N, O, Ne, Mg, and Fe. Hard X-ray coverage ($>$10 keV) will explore \textbf{coronal evolution and jet launching in local AGN}, while high-temporal resolution studies of \textbf{QPEs} will reveal the nature of unstable accretion. High-resolution observations of \textbf{jets in young stars} require a bandpass extending to $\sim$0.1--0.2 keV to resolve the interplay between jet and shock phenomena, establishing a critical physical link between stellar-scale jet launching and the powerful outflows seen in AGN. X-ray irradiation is a primary regulator of the chemical and physical architecture of \textbf{planet-forming disks} up to scales of $\sim$50 AU. To accurately quantify the energy deposition, sensitivity to the K-shell absorption edges of C, N, O, and Ne (0.28--0.87 keV) is vital, providing a necessary bridge between high-energy diagnostics and complementary radio/IR datasets. High-energy coverage up to 12 keV is also critical for \textbf{diagnosing powerful stellar flares,} which can reach plasma temperatures of 5--10 keV and disrupt exoplanet atmospheres. Sensitive spectra in this energy range will differentiate between single-temperature, multi-temperature, and non-thermal emission components. Finally, X-ray transit measurements offer a unique probe of the expanded upper \textbf{atmospheres of exoplanets}. Sensitivity in the soft X-ray band below 0.5 keV is fundamental for investigating these tenuous, high-altitude layers and provides critical context for HWO observations.
 
\subsection{Bandpass Analysis}
 
Selecting the energy bandpass is critical to achieving these science objectives. The Bandpass WG has defined limits based on both scientific necessity and technical feasibility.
 
\subsubsection{Upper Bandpass Limit: ~12 keV}
 
The extension of the energy range up to at least 12 keV is motivated by the need to enhance continuum anchoring and measure the signatures of accretion and outflows in local AGN and XRBs, facilitate the identification of heavily obscured spectra, characterize stellar flares, and ensure robust cross-calibration with concurrent hard X-ray missions. Going beyond this limit would likely require a more complex mission architecture, including a different focal length and more sophisticated multilayer coatings on the optics. The WG felt that science objectives requiring a higher upper limit on the bandpass would best be met by a separate mission.
 
\begin{itemize}
\item \textbf{Unveiling the First Black Holes:} A $\sim$12 keV upper limit anchors the coronal continuum and captures the full curvature of the Compton hump for sources at z $>$ 3, illuminating the telltale signatures of accreting supermassive black holes.
\item \textbf{Feeding and Feedback in Local AGN:} Maintaining as much usable effective area as possible at and beyond 10 keV enables important constraints on the above observables and also ensures that the mission is sensitive to ultra-fast outflows in the nearby universe, which typically manifest between $\sim$7-10 keV.
\item \textbf{Revealing the Physics of Transient Events:} High-energy coverage up to 12 keV is vital for characterizing the spectra of UFOs and accretion state changes in AGN, ULXs, and XRBs. It is also critical for diagnosing powerful stellar flares, which can reach plasma temperatures of 5--10 keV and disrupt exoplanet atmospheres. Sensitive spectra in this energy range will differentiate between single-temperature, multi-temperature, and non-thermal emission components.
\end{itemize}
 
\subsubsection{Lower Bandpass Limit: ~0.1 keV}
 
Extending the sensitivity down to 0.1 keV is essential for observing the soft X-ray signatures of hot gas within, around, and between galaxies, for characterizing accretion and outflows from supermassive black holes at high redshifts, for studying the chemical and physical architecture of planet-forming disks, and for understanding the physics at work in soft transients such as TDEs.
 
\begin{itemize}
\item \textbf{Feeding and Feedback at Cosmic Noon:} Crucial spectral lines such as O VII, O VIII, and C V, which trace AGN outflows, fall within the 0.1--0.2 keV range at cosmic noon (i.e., z $\sim$ 2-3), when AGN activity was at its peak in the Universe. Comparing the mass and energy of outflowing winds imparted into the ISM here vs. at later epochs will yield valuable insight into how AGN and their hosts have co-evolved over the last $\sim$2 Gyr.
\item \textbf{Cosmic Web and Large-Scale Structure}: Extending sensitivity to 0.1 keV will enable detection of high-redshift galaxy groups, the building blocks of cluster assembly. High-resolution soft X-ray spectroscopy of multiphase gas will trace thermal histories and mass assembly, providing critical constraints on structure growth and cosmological parameters ($\sigma_8$, $\Omega_{\rm m}$, dark energy, neutrino mass). These observations will also calibrate the non-gravitational feedback systematics that currently limit group-based cosmology.
\item \textbf{Young Stellar Jets:} X-ray-emitting jets from young stars probe the inner, most energetic regions of the accretion flow, where shocks and magnetic heating can raise material to millions of Kelvin. A bandpass extending down to 0.1--0.2 keV would enable reliable discrimination between different jet and shock components. This would unlock a new window into studying where and how the jets are heated, and how accretion, jet launching, and magnetic activity are coupled.
\end{itemize}
 
\subsection{Implications for Instrumentation}
 
To meet its scientific objectives and complement facilities in other wavelength regimes, the observatory must achieve high spectral resolving power (R $>$ 2000) across the recommended bandpass. A hybrid approach using both a microcalorimeter and gratings would be ideal in order to achieve this resolving power over the 0.1-12 keV range, as gratings are optimized for high resolving powers below $\sim$2 keV while microcalorimeters yield their highest resolving powers above $\sim$2 keV. The relative merits of both types of instruments are the purview of a different WG, but the bandpass recommended by this WG is compatible with such a hybrid approach. It would be particularly advantageous if the mission architecture allows both instruments to operate simultaneously.


\section{Microcalorimeter Working Group}
\sectionrule

An X-ray flagship calorimeter will revolutionize our view of the cosmos by doing what no other instrument can: transforming the invisible, superheated gases around black holes and exploding stars into spectacular 3D maps of the universe in motion. The microcalorimeter WG revisited the LXM architecture of the 2019/2020 CSR in light of post-2020 science (XRISM, IXPE, JWST, eROSITA, Einstein Probe, New Athena) and the maturation of microcalorimeter, cryogenic-readout, cooling, and large-area-filter technology. Its principal conclusions:

\begin{enumerate} 
\item \textbf{The LXM layout remains an excellent baseline:} Most of its assumptions have held up, and it is being actively advanced. The WG recommends adding well-targeted deltas rather than reopening it.

\item \textbf{The microcalorimeter instrument will provide transformative science across high-energy astrophysics:} \textit{Chandra}’s unprecedented angular resolution unlocked discoveries in the hot plasmas in galaxy clusters, relativistic jets from supermassive black holes, remnants of exploded stars, stellar winds, and planetary aurorae. Similarly, spatially resolved 1 eV energy resolution will answer questions that have been open for decades, as well as provoke new questions by opening new physical regimes. The power of high spectral resolution across these areas has been demonstrated recently by XRISM; Lynx will couple this capability with $\sim$100$\times$ higher angular resolution and $\sim$100$\times$ larger area, rendering the high-energy cosmos in three dimensions. 

\item \textbf{Line intensity mapping of extended hot diffuse gas is the strongest single motivation for the LXM:} the circumgalactic medium (CGM) of Milky Way-like galaxies, starburst outflows, galaxy cluster and group cores, outskirts, and filaments, the Galactic plane and center, and the Magellanic Clouds and M31. This is where the instrument has its largest qualitative advantage and where the community’s concurrent investment is strongest (Roman, Rubin, SKA, DSA-2000, and JWST).

\item \textbf{A graded-pixel focal plane is recommended:} sub-arcsec pixels in the central field for point-source and small-scale phenomena (e.g., galactic centers, AGN), and 2–5 arcsec pixels in an extended outer field for survey and faint-diffuse science. The realistic FoV target is $\sim$10–15 arcmin diameter; full 22$\times$22 arcmin coverage at half-arcsec everywhere is not achievable within credible readout, electronics, and cost constraints.

\item \textbf{Energy-resolution deltas should be modest and targeted:} Lowering main-array hydra multiplicity (the number of absorbers sharing a single readout sensor) plausibly improves it from 3 eV to $\sim$2 eV; a small inner sub-array near 1 eV up to $\sim$7 keV is feasible; the original $\sim$0.3 eV sub-array below 1 keV remains viable. A uniform $< 2$ eV across the full expanded FoV is not strongly motivated, and is in tension with the better-motivated FoV expansion.

\item \textbf{Extend usable energy range to $\sim$12 keV} via signal-reconstruction on TES devices, with degraded-resolution reach to $\sim$20 keV where the science requires it. Protect soft-X-ray reach to $\sim$0.3 keV (carbon edge) as a system-level requirement on the optical-blocking-filter (OBF) stack.

\item \textbf{Cost is the dominant constraint, not capability:} The Lynx 2030 microcalorimeter should be no more expensive than the original LXM, ideally less. 

\item \textbf{Passive cooling to 50 K is not recommended,} as it raises ground-program risk for an X-ray mission, and superconducting tunnel-junction detectors are not recommended as a primary technology (as they are count-rate-limited by carrier diffusion; $\sim$10 eV resolution at 7 keV).


\end{enumerate}

\subsection{The Lynx 2020 LXM Baseline}

All deltas stated in this document are referenced with respect to the CSR baseline. The LXM focal plane comprises three concentric sub-arrays plus a dropped extended-array option: 

\begin{table}[ht]
\centering
\renewcommand{\arraystretch}{1.15}
\setlength{\tabcolsep}{4pt}
\begin{tabular}{|C{3.6cm}|C{1.0cm}|C{1.65cm}|C{3.0cm}|C{5.0cm}|}
\hline
\rowcolor[HTML]{203864}
\color{white}\textbf{Sub-array} &
\color{white}\textbf{FoV} &
\color{white}\textbf{Pixel size} &
\color{white}\textbf{Energy resolution} &
\color{white}\textbf{Notes} \\
\hline

Main array &
$5'$ &
$1''$ &
3 eV ($\leq 7$ keV) &
$\sim$86,400 pixels; $5\times5$ hydras \\
\hline

Enhanced Main (EMA) &
$1'$ &
$0.5''$ &
$\sim$2 eV ($\leq 7$ keV) &
Higher-resolution center \\
\hline

Ultra-High-Res (UHR) &
$1'$ &
$\sim1''$ &
$\sim$0.3 eV ($<1$ keV) &
$\sim$3,600 pixels \\
\hline

Extended (option) &
$20'$ &
$2''$--$5''$ &
$\sim$1 eV ($\leq 2$ keV) &
Dropped in prior study \\
\hline
\end{tabular}
\end{table}

\subsection{How the Science Case Has Evolved Since Astro2020}

The original Lynx CSR organized its science into three themes — the Dawn of Black Holes, the Drivers of Galaxy Evolution, and the Energetic Side of Stellar Evolution. The single largest development since is the arrival of flight microcalorimeter spectroscopy: XRISM/Resolve has not merely confirmed CSR-era expectations but reframed several of them. For AGN, XRISM has established inner-accretion relativistic disk reflection as ubiquitous, resolving AGN relativistic outflow structure and power. In galaxy clusters, XRISM discovered dynamically active and super-metal-enriched regions around supermassive black holes and complicated dynamical flows of hot plasma of merging clusters. For supernova remnants (SNRs), XRISM revealed velocity structures along the line of sight to provide a 3D map of the supernova ejecta and detected the odd-Z elements Cl and P for the first time in X-rays to constrain nucleosynthesis models. IXPE, JWST, and the eROSITA all-sky survey have likewise reshaped the landscape, and NewAthena XIFU ($\sim$8 angular resolution, late 2030s) will probe these same regimes with higher angular and spectral resolution than XRISM/Resolve. The net effect is to shift the center of gravity of the Lynx 2030 microcalorimeter case toward the regime where it retains a unique, qualitative advantage — hot diffuse gas in emission — and to re-cast the AGN and SNR cases as the cosmological extension and routine application of techniques whose existence proofs are now largely in hand.

\subsubsection{Drivers of Galaxy Evolution --- hot diffuse gas in emission
}

\begin{figure}[h]
\centering
\includegraphics[width=0.8
\textwidth]{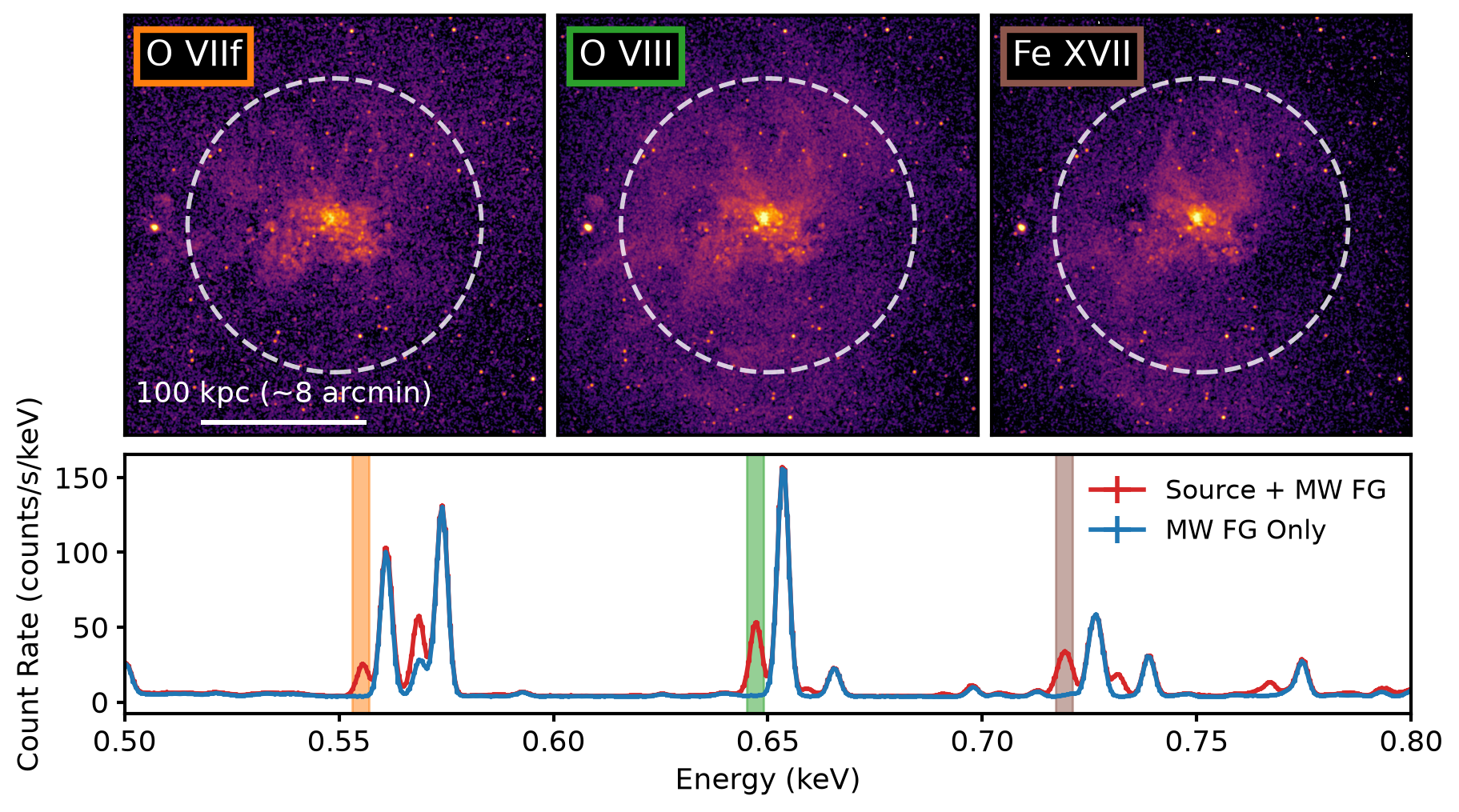}
\caption{Simulated LXM observations of the CGM of a Milky Way-mass galaxy at $z = 0.01$ from the TNG 50 simulation \cite{TNG50}, assuming the addition of a 20\arcmin\ extended array with 2\arcsec\ pixels. The microcalorimeter resolution enables separation of the redshifted O VIIf, O VIII, and Fe XVII lines of the external galaxy from the Milky Way (MW) foreground, enabling maps of different gas phases (top panels) to be made from the narrow bands from the extracted spectrum from the circular region (bottom panel), as shown. The MW foreground (blue line) largely obscures the CGM from the extragalactic source (red line), highlighting the need for high spectral resolution for this science.}
\label{fig:cgm_lxm_example}
\end{figure}

This is the WG’s flagship case, and it is where the science case has shifted most decisively since the CSR. The CSR leaned on absorption-line studies of the diffuse universe;  however, using absorption spectroscopy of extragalactic diffuse gas as a primary driver is risky due to the small spatial sampling by design, limited availability of X-ray bright quasars (compared to those in UV/optical) and their alignment with intervening halos, and a restricted redshift window that is free from z = 0 transitions in the multiphase ISM/CGM of the Milky Way. These considerations have led this WG to focus on emission-line intensity mapping. A few-eV microcalorimeter, integrating in individual emission lines, suppresses the soft-X-ray foreground by a factor of order 30 relative to a CCD, enabling line intensity mapping of key systems on resolved scales. Concrete cases include:

\begin{itemize}

\item \textbf{Hot halos of ordinary galaxies:} The role of the multiphase CGM is instrumental in the disk-halo cycle of metals, baryons, and energy, and the well-observed cooler phases of the CGM are missing most of the key answers. It is essential to survey the hot ($\gtrsim 10^6$ K) CGM of a large galaxy population to address the intrinsic halo-to-halo scatter and the huge disagreement across simulations due to diverse feedback prescriptions \cite{Schellenberger2024,Silich2025}. Unambiguous detection of the signal requires a spectroscopic segregation from the dominant and complex z = 0 Galactic foreground (see Figure \ref{fig:cgm_lxm_example}), which is beyond the capabilities of current instruments, and sets the resolution requirement for LXM ($\sim$3 eV at typical cosmological redshifts). Lynx will go beyond just the detection of OVII and OVIII lines (along with C, N, Ne, Mg, and Fe L-shell tracers) and unveil detailed maps of thermal, kinetic, and chemical properties at a $\sim$1 kpc scale of the inner ($\sim$50 kpc) CGM of z > 0.05 edge-on galaxies \cite{ZuHone2024}.  

\item \textbf{Starburst-driven outflows} are demographically more important than AGN winds for the baryon and metal cycle in star-forming galaxies, yet the only existing high-resolution measurements come from the small XRISM sample \cite{XRISM_M82_2026}. They emit primarily in the soft band where the foreground-rejection advantage is greatest. Velocity-resolved line maps can separate the X-ray-emitting hot phases from the cool and warm phases and measure kinetic power and mass-loading directly; a single pointing captures the wind, its deceleration into the halo, and the halo response together — exactly the chain simulators most need anchored. In current simulations, starburst and AGN outflows are assumed to follow almost identical physics, with numerical values hand-tuned to match limited observations of starbursts. With simultaneous 2-D constraints on kinetic energy, thermal energy, and the chemical composition of starburst outflows and their interaction with the surrounding CGM, Lynx will test this assumption.

\item \textbf{AGN feedback in galaxy cluster and group cores, and host galaxies:} One of Chandra’s most important legacies has been to map in exquisite detail the central regions of galaxy clusters and groups, detailing the powerful effects of central AGN feedback \cite{Blanton11,Fabian2011,2015ApJ...805..112R,Fabbiano2022}. Combined with radio observations that trace cosmic rays and magnetic fields, and optical observations that probe cool filaments condensing out of the hot phase, a physical picture of how AGN feedback works is emerging, but the key question–how does the energy injected into the ICM atmosphere heat the gas, is still unresolved. XRISM/Resolve has provided the first direct measurements of the kinetic power of cluster AGNs \cite{XRISMPerseus2026,XRISMVirgo2026,XRISMAbell2029}, but lacks the angular resolution to map the kinematic energy as a function of scale. Given the excellent photon statistics available in cluster cores, Lynx will map the thermal and kinematic properties of the regions surrounding AGN at the scales required to test simulations and determine the dynamical coupling with other gas phases.

\item \textbf{Cluster outskirts and intercluster filaments:} Spectroscopy beyond virialized regions and into the cosmic-web filaments feeding clusters accesses temperature, metallicity, and bulk-flow kinematics at the cluster–filament interface \cite{Zhang2024}. The surface brightness is intrinsically low; while XRISM/Resolve can perform ICM spectroscopy up to $R_{\rm 2500c}$ \cite{XRISMPerseus2026,Sarkar2026}, only Lynx 2030’s combination of area, spectral resolution, and per-line foreground rejection makes the regime beyond $R_{\rm 200c}$ accessible at all, and only $\sim$1 eV spectral resolution will be able to resolve individual lines from various phases of the gas.

\item \textbf{Resolving the hot ISM at the physically meaningful scale:} The ISM is the most crucial component of the metal, baryon, and energy cycle within the galactic disks. The physics driving the properties of ISM operates at a critical sub-galactic ($\sim$100 pc) scale and is expected to vary across the disk. These scales are well-constrained in all other ISM phases except the hot phase. Lynx will spatially separate the sub-galactic hot ISM spectra from X-ray binaries of $z < 0.05$ face-on galaxies, and constrain the temperature, ionization mechanism, equilibrium condition, and chemical composition, which are beyond the scope of any current or upcoming missions.

\item \textbf{Absorption line studies:} Stacked AGN sightlines deliver O VII/O VIII absorption by intervening diffuse gas from the WHIM and the CGM of external galaxies \cite{Bogdan23} (cross-correlated with SZ- and FRB-based gas measurements) and the Milky Way as a side product. These will be natural byproducts of AGN science (see below) and deep exposures on faint, diffuse objects, without investing additional resources. 

\end{itemize}

\subsubsection{Dawn of Black Holes --- inner accretion flow and AGN outflows}

\textbf{Inner accretion flow and the cosmological evolution of spin.} Building on previous CCD detections and grating measurements, XRISM/Resolve has shown that broad, relativistically broadened Fe K$\alpha$ lines, produced as X-rays are reflected from the innermost radii of accretion disks, is essentially ubiquitous in unobscured AGN once microcalorimeter resolution separates the narrow lines and absorbers from the continuum \cite{Brenneman2025,XRISM_NGC4151_2024,Kammoun2025,Li_NGC3783}. Measurements of relativistically broadened Fe K$\alpha$ lines enable studies of the inner accretion disk and robust measurements of black hole spin \cite{reynolds+2021}. The profile of the relativistically broadened emission line is remarkably consistent across six orders of mass, between these AGN and the black hole X-ray binary Cygnus X-1, showing that the broad Fe K$\alpha$ line is produced by the same fundamental dynamics of material in orbit around a spinning black hole, and is hard to reconcile with alternative explanations of the broad line origin\,\cite{Brenneman2025,Wilkins2026,Draghis2025}. The Lynx 2030 case is the cosmological extension of X-ray reflection studies of black holes and the innermost regions of their accretion flows. Spin encodes growth history (coherent accretion spins up, mergers and chaotic accretion spin down), probed through the spin distribution of black holes as a function of mass \cite{berti_volonteri,bustamante_springel,piotrowska+2024}. While the current state-of-the-art measures the spin-mass relationship across AGN in the local Universe \cite{reynolds+2021,Mallick2022,Mallick2025}, Lynx 2030 will measure the spin–mass distribution versus redshift to weigh accretion- versus merger-driven growth across cosmic time \cite{Pacucci_2020}. With $\sim$100$\times$ XRISM’s area at sub-arcsec focal-plane resolution, Lynx 2030 extends spin measurements to z $\sim$ 3 to within a precision of better than 8\% and toward z $\sim$ 5-6 in the fastest spinners. The spectral resolution of the microcalorimeter is critical, however, to ensure that narrow components of the Fe K$\alpha$ line, in addition to absorption lines in the Fe K band, can be unambiguously separated from the underlying broad line profile produced by reflection from the inner accretion disk. With this spectral resolution, the effective area and angular resolution of Lynx 2030 will provide the required signal-to-noise ratio at these high redshifts for precision black hole spin measurements.

\textbf{AGN outflows.} XRISM decisively confirmed the presence of ultra-fast outflows: hot, relativistic winds in AGN with enough power to affect their host galaxies. We now know that these likely look different in high-Eddington sources, like PDS~456 and PG~1211+143 \cite{XRISM_PDS456_2025,Xu2025,Mizumoto2026}, where the UFO has been resolved into multiple distinct kinematic zones, compared to a single broad absorption feature in low-Eddington AGN like MCG-6-30-15 and NGC~3783 \cite{Brenneman2025, Mehdipour2025, Gu2025}. Thanks to its higher effective area and spectral resolution, Lynx 2030 will tightly constrain the mass, momentum, and energy carried by AGN winds, improving measurements of their kinetic power and feedback efficiency. Its calorimeter line profiles will also test wind-launching mechanisms \citep{Gandhi2022, Fukumura2022} and extend high-resolution wind studies beyond the local Universe, building samples at redshifts more relevant for AGN feedback and galaxy evolution, where current observations are still limited by CCD resolution \citep{Matzeu2023}. In addition, Lynx’s sub-arcsecond imaging will resolve the spatially extended wind component in emission, opening a new view of AGN feedback on host-galaxy scales. These cases are major beneficiaries of the focal-plane design, which delivers them automatically at the center. The same sub-arcsec central pixels also enable resolved AGN-feedback mapping in nearby host galaxies, where Chandra has revealed a spatially and spectrally complex picture: ionization cones, jet-termination shocks, cross-cone outflows, and Fe K$\alpha$ fluorescence from circumnuclear molecular gas on 30–100 pc scales. Pairing that sub-arcsec resolution with a microcalorimeter and a collecting area that is $\sim$30–50$\times$ Chandra’s turns today’s descriptive morphology into quantitative kinematics, measuring outflow and shock velocities directly across the $\sim$20 nearby Compton-thick AGN already mapped with Chandra \cite{Fabbiano2022}.

\subsubsection{Energetic Side of Stellar Evolution --- SNe, SNRs, stars and exoplanets}

Although the New Athena XIFU, with $\sim$8” angular resolution, will address several questions related to core-collapse and Type Ia supernova physics, Lynx will be uniquely positioned to address deeper and more fundamental questions related to the cradle-to-grave evolution of massive stars, the formation of trace elements \cite{XRISM2026CasA,Boccioli2026}, and the basic problem of energy partitioning between ions and electrons in strong shocks \cite{Ghavamian2013}. Addressing these issues requires a large FoV and spectral resolution superior to New Athena. For instance, with 30-50$\times$ the effective area of \textit{Chandra}, follow-up observations of historical and recent supernovae with Lynx will probe the mass loss histories of supernova progenitors \cite{Patnaude2017} out to $\sim$20 Mpc. Alongside these studies of recent supernovae, arcsecond spatial-resolution imaging of their host galaxies will enable SNR population studies far beyond the Local Group \cite{Leonidaki2010}, addressing questions such as the ``missing supernova remnant'' problem. Additionally, studies of Galactic and Magellanic Cloud remnants will address questions related to the synthesis of Fe-group elements in core-collapse and Ia SNe \cite{Sato2026}. Observations of remnants such as G292 will also probe element formation during the explosion due to neutrino anisotropies. Finally, per-species ion temperature measurements in intermediate-velocity collisionless shocks (Tycho, Kepler, RCW 86, Cas A), the regime between fast SN 1987A and slow Cygnus-Loop shocks that is currently unconstrained. In addition to studying the end points in stellar evolution, connections will be made to the Astro2020 ``stars and planets in context'' theme and the exoplanet community through Lynx observations of stellar nurseries such as the Orion Nebula \cite{Wolk2005}, and the interactions between exoplanets and their hosts. 

\subsubsection{Synergies with the 2030s/2040s landscape}

The recommended architecture is well-matched to concurrent facilities: Roman and Rubin (matched wide-field optical/NIR companions), HWO (probing the lower ionization states of metals in UV/optical), SKA/ngVLA/DSA-2000 (AGN feedback, cluster radio-mode physics, FRB sightlines for the WHIM), JWST (high-$z$ baryon-cycle context), mm/sub-mm telescopes like CCAT/SO/AtLAST (probing hot gas temperatures and velocities via the SZ effect), and LISA (SMBH-merger counterparts via the sub-arcsec inner array). The relationship to New Athena is complementary — New Athena’s XIFU leads at higher energies, while the Lynx 2030 microcalorimeter’s distinctive contributions are sub-arcsec calorimeter pixels, soft-band reach down to $\sim$0.3 keV, and wide-field spectroscopy at few-eV resolution. These synergies and Lynx 2030 microcalorimeter capabilities open a number of impactful TDAMM opportunities, similar to what was achieved in the TDAMM field with \textit{Chandra}.

\subsection{Architectural Deltas --- Answering the Terms of Reference Framing Question}
\label{subsec:deltas}

\subsubsection{Larger field of view --- the principal delta}

FoV expansion is the WG’s main architectural change. The endpoint is set by readout-electronics scaling, not detector fabrication: a factor-30-to-60 channel increase ($\sim$20-30\arcmin\ width at 0.5\arcsec) is not credible because warm-electronics power, mass, and volume become severe. Rather, the WG recommends an outer wide-field array of $\sim$10–15 arcmin diameter with 2–5 arcsec pixels. The principal cosmic ecosystems science objectives can still be achieved with such an array; for example, at a distance of $\sim$60 Mpc, a 10-arcmin field captures most of a galaxy’s inner CGM. The 10–15 arcmin-wide array proposal aims to provide Lynx with wide-field grasp and sub-arcsec PSF together, at calorimeter resolution with soft-band reach, a combination no other instrument provides.

\subsubsection{Energy resolution across the FoV}

\textbf{Whether uniform < 2 eV across a full and extended FoV strengthens the case:} the WG’s answer, after investigating multiple science cases, is not strongly in favor. The cases that genuinely need < 2 eV are concentrated at the center (resolved AGN spectroscopy, inner-disk reflection, outflow-zone separation, hot gas flows and turbulence in the centers of galaxies and galaxy clusters), and will fit within the main array. The wide-field diffuse-gas cases need few-eV resolution but are limited at the periphery by surface brightness and photons per spaxel, not by spectral resolution. Therefore, uniform < 2 eV across an expanded FoV multiplies channel count without proportional science return. The WG therefore recommends spending resolution where the science benefits:

\begin{itemize}
\item Main array (5$\times$5\arcmin, 1$''$): improve $3$ eV $\rightarrow$ 2 eV by reducing hydra multiplicity; useful to $\sim$12 keV.
\item Enhanced Main array (1$\times$1\arcmin, $\sim$0.5$''$): 2 eV up to 7 keV, using small-absorber hydras prototyped at GSFC.
\item UHR sub-array ($\sim$1$\times$1\arcmin): $\sim$0.3--0.4 eV below 1 keV.
\item Outer wide-field array ($\sim$10--15\arcmin, 2--5$''$ pixels): 2.5 eV up to $\sim$2 keV, primarily for large, diffuse, and faint objects.
\item Optional 1-eV inner sub-array ($\sim$1$\times$1\arcmin, $\sim$1\arcsec): $\sim$1 eV up to 6--7 keV for compact-object spectroscopy.
\end{itemize}

\subsubsection{Broader bandpass}

\textbf{Above 7 keV:} signal reconstruction on TES devices gives useful spectroscopy ($<10$ eV) to $\sim$12 keV (recommended target) and, with development, $\sim$20 keV, benefiting the AGN continuum, Fe K$\beta$, cluster thermodynamics, and ion-temperature work, at essentially zero mass/cost impact. \textbf{Below 0.5 keV:} soft-band reach to the carbon edge ($\sim$0.28 keV) is a system-level requirement — the response, OBF stack, and operating temperature must all be consistent with it to enable high-redshift science. For example, at $z \sim 10$, rest-frame Fe lines shift to 0.3–0.6 keV, the main window onto Compton-thick early growth of black holes. 

\subsubsection{Pixel scale}

A microcalorimeter is an IFU: spectroscopy needs counts per spaxel, so very small pixels buy mapping only where the photon budget is rich. This favors a small inner sub-array of very-small pixels rather than wide-field very-small pixels. The WG recommends a graded scale — a small High-Resolution Inner Array ($\sim$20–60") at $\sim$0.5", the 0.5–1" main array, and 2–5" in the outer field. This standardization is supported by dither and sub-pixel reconstruction: for a point source, the reconstructed image width combines the optic PSF and the pixel-quantization term in quadrature, so for a sub-arcsec optic the gain from finer pixels is small. With a 0.5"-HPD PSF, going from 0.5" to 0.33" pixels narrows the reconstructed point-source width only from $\sim$0.26" to $\sim$0.23" ($\sim$10\%). This is too little to justify a finer 0.33" array, thus the inner array is standardized at 0.5", with the optic PSF the dominant driver of resolution (and the leading cost and risk).

\subsection{Technology Maturity and Cost}

\textbf{Five post-2020 technology shifts underpin the recommendation.} Most consequentially, (1) relocating the cryogenic HEMT low-noise amplifiers from the 4 K stage to the $\sim$20 K cryocooler stage effectively eliminates the 4~K cooling-power bottleneck that previously constrained the readout channel count (the single biggest enabler of a wider graded field). At 20 K, the robust cooling capacity can comfortably support 64 readout channels, leading to $\sim$4$\times$ more channels; (2) a single 4--4.5 K cryocooler now demonstrated to high TRL; (3) microstrip flex wiring replacing coax, resulting in more compactness and lower heat loads; (4) a 40 mK bath via continuous-ADR instead of 50 mK; (5) digital readout: either a near-term flight-ready Kintex UltraScale + discrete GSPS converters baseline, or AMD Versal RF (space-qual announced Nov 2025, parts expected 2029).

\textbf{Cost.} Every recent X-ray flagship has been judged scientifically excellent; the science case is not the binding constraint. The Lynx 2020 LXM was the most expensive single instrument in that concept; the Lynx 2030 microcalorimeter should be no more expensive, ideally less, achieved by exploiting technology maturity and the larger mass/power envelopes ($\sim$15 t, $\sim$100 kW) of new launch vehicles rather than expanding scope. These trades are not unique to Lynx: NewAthena’s reformulation similarly balanced the field of view and cooling architecture against a fixed cost envelope, settling on a 4' X-IFU field and passive 50 K cooling. The parallel underscores that wide-field calorimetry is cost-driven, and that the cooling chain is a leading cost-and-risk lever — which is why the WG emphasizes retiring that risk early.

\textbf{Front-loaded-investment pathway.} A distinct option is to invest heavily in LXM technology before the costly mission ``marching army'' begins charging — sustained technology funding a factor 3–4 above recent levels ($\gtrsim$ \$500M over 5–10 years for the full mission portfolio), retiring risk on the key system-level demonstrations (a 100,000+ pixel array with partially-instrumented readout, the post-2020 readout chain at full channel count, large tiled OBFs with contamination mitigation and an aperture design balancing thermal loads and IR shot noise, full-$\mu$MUX digital readout). Such demonstrations at the LXM scale have not yet been performed — the CSR assumptions were sound at the device level but unvalidated as a system. If achieved, the mission could be proposed as a higher-risk protoflight on a ~5-year build within a $\sim$\$3B cap, with the EM-vs-protoflight risk/savings trade left to the concept-study team.

\section{New Capabilities Working Group}
\sectionrule

The Lynx X-ray Observatory concept presented to the Astro2020 Decadal Survey defined a flagship-class soft X-ray telescope built around three scientific pillars: black hole seeds at cosmic dawn, the drivers of galaxy evolution, and the energetic side of stellar evolution. The science goals associated with these pillars were enabled by a subarcsecond-class optic with $\sim 2 \, \rm m^2$ of effective area at $1$ keV, a high-density imaging detector, an ultra-high-resolution microcalorimeter array, and a dispersive grating spectrometer. In the half-decade since Astro2020, the Lynx science case has, in many ways, been strengthened. For example, JWST's discovery of the ``Little Red Dots'' (LRDs), now widely interpreted as X-ray-weak, accreting massive black holes in the hearts of infant galaxies, has sharpened the demand for the deep, high-resolution X-ray imaging spectroscopy that only a Lynx-class facility can deliver.
This document tracks how the key enabling technologies of the Lynx design have advanced in their Technology Readiness Level (TRL) since 2020, driven largely by six years of continued technology development partly coupled to a cohort of Probe- and Explorer-class studies, including STAR-X and AXIS (high angular resolution imaging enabled by silicon meta-shell optics), LEM (microcalorimeters), Arcus (critical-angle transmission gratings), and HEX-P (hard X-ray response), together with the on-orbit successes of IXPE and XRISM. It then follows the working group's lead in surveying new capabilities—most prominently high-resolution X-ray polarimetry—that were not part of the 2020 baseline but are now sufficiently mature to be considered for mating with a Lynx-class optic. In addition, new launch and spacecraft capabilities may open doors beyond the original Lynx concept.

\subsection{Introduction and Motivation}

Following the recommendations of the Astro2020 Decadal Survey and navigating the shifted astrophysics landscape of the mid-2020s, it is timely to revisit the Lynx reference design to identify the transformative science that can be unlocked by incorporating mature, next-generation detector technologies. Recent discoveries from IXPE, XRISM, and JWST have underscored the necessity of capabilities such as high-resolution X-ray polarimetry and rapid-readout timing, while reaffirming the centrality of the optic and the microcalorimeter that anchored the original concept.

The motivation has a distinctly post-JWST character. The Little Red Dots are extremely compact ($< 100$ pc \cite{Baggen_2023}), red-continuum sources with broad Balmer lines—signatures of dust-obscured active galactic nuclei (AGN) that are remarkably, even anomalously, weak in X-rays, with non-detections persisting even in deep stacks \cite{Maiolino_2024_Xray, Ananna_2024, Yue_2024_Xray}. Whether this reflects intrinsic super-Eddington accretion \cite{Pacucci_Narayan_2024, Jones_2025}, Compton-thick obscuration \cite{Maiolino_2024_Xray}, or reprocessing in dense ionized cocoons, the question is fundamentally an X-ray one. Distinguishing these scenarios for individual sources at $z \gtrsim 5$ requires the combination of large effective area, arcsecond imaging to beat down confusion and background, and the spectral resolution to characterize absorption and ionized winds—precisely the parameter space Lynx was designed to own.

By combining the revolutionary effective area and angular resolution of a Lynx-like optic with modern detectors—advanced Gas Pixel Detectors (GPDs), rapidly maturing microcalorimeter arrays, and Critical-Angle Transmission (CAT) gratings—the observatory can access compelling science cases that are unavailable to the original baseline instrument suite. The new launch and spacecraft capabilities may open doors beyond the original Lynx concept. The remainder of this section is organized in two parts: Part I (\S4.2) documents the maturation of the baseline Lynx technologies; Part II (\S4.3) surveys new capabilities. \S4.4 addresses mission architecture and implementation strategy, and \S4.5 summarizes the TRL landscape.

\subsection{Part I — Maturation of the Baseline Lynx Technologies}

The 2020 Lynx concept rested on four enabling technologies, each with a defined development roadmap and a target TRL of 6 prior to mission start \cite{Lynx_CSR}. Crucially, the intervening Probe and Explorer studies—and two operating microcalorimeter/polarimetry missions—have served as de facto pathfinders, advancing the very components Lynx requires. We treat each in turn.

\subsubsection{X-ray Optics: Silicon Meta-Shell, Full-Shell, and Glass}
The optic remains the pacing technology for any Lynx-class flagship, and maintaining a (quasi) monolithic optic design will depend on the interplay between key scientific drivers and the realistic advancement of optic technology. Three architectures are being actively matured:

\subsubsection*{Silicon Meta-Shell Optics (SMO/SMSO — GSFC)}
The Lynx baseline optic was the silicon meta-shell optic: more than 37,000 precision-polished, 0.5 mm-thick monocrystalline silicon segments, layered onto structural shells, combining the fabrication advantages of segmented optics with the alignment advantages of full shells. Since 2020, two Explorer-class Phase A studies have advanced this technology. STAR-X (a MIDEX concept) adopted the silicon meta-shell mirror assembly, and full-illumination X-ray tests of an uncoated single mirror-segment pair have demonstrated a 2.2-arcsecond half-power diameter (HPD) at the segment-pair level.
AXIS, the Advanced X-ray Imaging Satellite (a Probe-scale concept, now not moving forward), proposed applying the same technology at a substantially larger scale: a 9 m focal length, 1.8 m outer-diameter assembly of 298 shells grouped into six meta-shells ($\approx 2,700$ segments each), iridium-coated, with a design goal of $> 7,700 \, \rm cm^2$ at 1 keV and $\approx 1,600\, \rm cm^2$ at 6 keV and a near-constant 1.6\arcsec HPD across the field—roughly $15\times$ Chandra's area at 0.5 keV \cite{AXIS}. These figures are design targets rather than demonstrated on-orbit or full-assembly performance. To date, the strongest experimental results have been achieved at the segment-pair level; as the working group notes, individual silicon meta-shell segments can meet Lynx requirements, but integration introduces serious challenges even at small scales. The scalable integration of thousands of mirror segments into aligned meta-shells remains the labor-intensive, risk-bearing step and has not yet been demonstrated at the required scale or performance. Recent roadmap milestones target X-ray verification of multi-module meta-shells through environmental testing—the essential bridge from the current segment-pair demonstrations toward the integrated-assembly performance a large mission demands.

\subsubsection*{Full-Shell Optics (MSFC)}
An alternative path electroforms full nickel-alloy shells from mandrels polished to sub-arcsecond levels, reliably achieving FWHM $< 2$\arcsec and offering high-throughput, simplified alignment. Ongoing efforts seek to mitigate the stochastic spatial-frequency errors introduced upon shell release—the dominant error term for this architecture. This approach remains viable, particularly where simplified integration is prioritized over the absolute thinnest mass budget.

\subsubsection*{Glass Optics (SAO)}
Directly polished, heat-formed, or slumped-glass optics carry a strong heritage from Chandra, combining a highly controlled surface polish with good coating compatibility and low-complexity integration. This heritage-based approach is highly viable, particularly if new launch vehicles mitigate previous mass-related concerns (\S4.4). SAO is currently working with OPTIMAX to fabricate a ULE heat-formed segment that incorporates both the P and S shells into a single monolithic segment (thereby greatly reducing the total number of segments and the need to align pairs).

\subsubsection*{Coatings}
Low-stress multilayer coatings are an area of active, rapid development, essential for preventing deformation in segmented designs and thereby ensuring broader energy response at a given focal length. Multilayer coatings also enhance hard-energy response within a desired waveband, enabling larger graze angles and field of view—technology that matured directly through the hard X-ray missions discussed in \S4.3.2.

\subsubsection{X-ray Microcalorimeters}

X-ray microcalorimeters remain the unequivocal priority for high-energy-resolution imaging spectroscopy, and recent developments have made their inclusion on a Lynx-class mission even more compelling. The Lynx X-ray Microcalorimeter (LXM) baseline called for an unprecedented $>$ 100,000-pixel transition-edge-sensor (TES) array behind the 0.5\arcsec, $\sim 2 \, \rm m^2$ optic, with a main array of 1\arcsec pixels over a 5\arcmin field, an enhanced central array of 0.5\arcsec pixels, and $\sim 3$ eV resolution \cite{Bandler_2019}. The technology is advancing robustly toward TRL-5 with respect to the original Lynx microcalorimeter requirements.

Two developments since 2020 are decisive. First, XRISM/Resolve (launched September 2023) has now flight-validated the core measurement, delivering $\sim$4.5–5 eV FWHM across 0.3–12 keV from a 36-pixel array—the on-orbit demonstration that closed-cycle, space-qualified microcalorimetry works as designed. Second, the LEM Probe study has advanced the large-format, fine-resolution array directly relevant to LXM: a $\sim$14,000-pixel TES array spanning a wide field, with an inner 7\arcmin×7\arcmin sub-array of single-TES pixels delivering 1 eV resolution and the outer ``hydra'' pixels ($2\times2$ absorbers per TES) reaching $\sim 2$ eV below 2 keV. The LEM array and its readout leverage mature heritage (approaching TRL-5, with aspects assessed consistent with TRL-6 in independent NASA/GSFC reviews).

Crucially, new developmental pathways have emerged that could make the instrument both easier and more cost-effective to produce. Full-scale Lynx-2020 detector arrays are already in active development, demonstrating that manufacturability is well in hand and not a primary mission risk compared to the telescope optics. Readout technologies—microwave-multiplexed SQUIDs in particular—are advancing rapidly, greatly aided by parallel investments and breakthroughs in quantum computing. These advancements make wider-field-of-view array concepts highly realistic to consider, e.g., combining high-resolution small pixels (0.33\arcsec) in the center with larger pixels (0.5\arcsec–2\arcsec) in the outer regions to expand the field from the original 5\arcmin to a transformative 10–20\arcmin, particularly optimized for soft X-rays.

\subsubsection{Critical-Angle Transmission (CAT) Gratings}
A significant technical advance since the original Lynx concept study is the maturation of Critical-Angle Transmission (CAT) gratings, driven principally by the Arcus Probe/Explorer studies. The original Lynx Design Reference Mission carried a grating resolving-power requirement of $R \approx 5,000–7,500$ \cite{Lynx_CSR} However, recent experimental demonstrations have proven that CAT gratings themselves can achieve $R > 10,000$: coaligned Arcus prototype facets illuminated by a silicon-pore optic at the PANTER facility reached an effective resolving power of $\approx1.3 \times 10^4$ ($3\sigma$), a factor of $\sim 2–4$ above the Arcus requirement \cite{Heilmann_2016, Heilmann_2022}. The gratings are therefore no longer the limiting optical element.

To realize a Lynx-like CAT X-ray Grating Spectrometer (CATXGS) at $R \approx 10,000$, the entire optical design—especially the focusing mirror array—must support this resolution, likely necessitating a tightening of the 0.5\arcsec angular-resolution budget and detailed ray-tracing to define tolerances and the R error budget. Additionally, new theoretical grating designs suggest it may be possible to extend the useful high-resolving-power bandpass toward 4 keV, and potentially up to 6 keV \cite{Heilmann_2016}

\subsubsection{High-Definition X-ray Imaging Detectors}
The Lynx baseline also carried a wide-field High-Definition X-ray Imager (HDXI) with active-pixel CMOS or monolithic CMOS sensors, finely sampling the PSF. This detector class has matured steadily through the same Probe studies—AXIS, for example, baselines a high-speed CMOS camera with active ground-calibration programs now underway—so that fast, low-noise, fine-pixel imagers oversampling an arcsecond PSF are no longer a development risk relative to the optic \cite{AXIS}.

\subsubsection{Pathfinder-to-Lynx Technology Mapping}

Table \ref{tab:lynx-technology-trl} below summarizes how each post-2020 study has advanced the corresponding Lynx baseline technology.

\begin{table}[H]
\centering
\renewcommand{\arraystretch}{1.25}
\begin{tabular}{|C{3.8cm}|C{3.0cm}|C{4.2cm}|C{2.6cm}|}
\hline
\rowcolor[HTML]{203864}
\color{white}\textbf{Lynx technology} &
\color{white}\textbf{Studies since 2020} &
\color{white}\textbf{Key demonstrated advance} &
\color{white}\textbf{Indicative TRL} \\
\hline

\textbf{Silicon meta-shell optics} &
STAR-X, AXIS Phase A studies &
$2.2''$ HPD uncoated segment pair; AXIS modules at $1.6''$ over field; $>7{,}700~\mathrm{cm}^{2}$ @ 1 keV &
3 $\rightarrow$ 4 (segment/module) \\
\hline

\textbf{Microcalorimeter array (LXM)} &
LEM concept study; XRISM/Resolve (flight mission) &
14k-pixel TES array, 1--2 eV; flight 4.5--5 eV (Resolve); $\mu$mux readout &
$\approx$4 for a Lynx-class microcalorimeter, with high ADD \\
\hline

\textbf{CAT grating spectrometer} &
Arcus concept study &
$R \approx 1.3\times10^{4}$ measured (PANTER); 2--4$\times$ requirement &
$\approx$5 (grating element) \\
\hline

\textbf{Wide-field CMOS imager (HDXI)} &
AXIS, STAR-X Phase A studies &
High-speed fine-pixel CMOS cameras in ground calibration &
$\approx$5 \\
\hline

\textbf{Hard-energy multilayer coatings} &
HEX-P concept study &
Pt/C, W/Si depth-graded multilayers; focusing to 80--200 keV &
$\approx$5 (NuSTAR heritage) \\
\hline
\end{tabular}
\caption{TRL values are indicative, drawn from the cited study reports and roadmaps; component vs. system-level TRL differs and should be read with the integration caveats in the text. ``ADD'' stands for ``advancement degree of difficulty'', i.e., how difficult it is for the technology to step to the next TRL level. }
\label{tab:lynx-technology-trl}
\end{table}

\subsection{Part II — New Capabilities Beyond the 2020 Baseline}
Beyond maturing its baseline, the landscape since 2020 has produced new detector capabilities mature enough to consider mating to a Lynx-class optic—capabilities the original concept did not include. Foremost among these is high-resolution X-ray polarimetry, demonstrated on-orbit by IXPE and now being scaled by eXTP; we also address extending focusing to hard X-rays (HEX-P) and wider-field/time-domain detector concepts.

\subsubsection{High-Resolution X-ray Polarimetry}
The recent success of IXPE has unequivocally demonstrated the power of X-ray polarimetry for constraining astrophysical models. IXPE's three telescopes, launched in December 2021, established Gas Pixel Detectors in flight across 2–8 keV, reaching a minimum detectable polarization (MDP$_{99}$) below $\sim 5.5\%$ for a 0.5 mCrab source in 10 days. However, current observations are severely hampered by limited effective area and angular resolution, restricting studies to a handful of marginally resolved sources measured with modest precision over megasecond exposures.

By mounting enhanced GPDs behind a Lynx-class arcsecond-resolution, square-meter-class optic, the observatory transitions from sensitivity-limited, time- and space-averaged ensemble measurements to rich, dynamic studies of at least thousands of wide-ranging targets. A next-generation polarimeter on a large-area observatory would open entirely new frontiers across Galactic and extragalactic scales; the phenomena are highly dynamic, and a Lynx-class effective area enables access to polarization changes on timescales currently unavailable.

\subsubsection*{Science drivers for Lynx-class Polarimetry}
\begin{itemize}
    \item \textbf{Galactic accelerators and cosmic rays:} Mapping field geometry and turbulence in supernova remnants probes the diffusive shock acceleration of cosmic rays. The TeV sky is dominated by relativistic outflows such as pulsar wind nebulae and the SS 433 jets; arcsecond-resolution polarization maps extract their properties without the spatial averaging that dilutes polarization signatures.
    \item \textbf{Black hole and neutron star geometries:} For point sources, X-ray polarization uniquely unlocks the unresolved geometries of accretion disks, jets, and coronae; for neutron-star binaries, the angle between the spin axis and the magnetic dipole axis can be determined.
    \item \textbf{Fundamental physics and magnetars:} Mapping the changing field geometries of magnetized neutron stars measures free precession, magnetic-field evolution, and magnetosphere QED effects. Pushing GPD measurements to 30–50 keV opens a window onto the cyclotron lines of highly magnetized binary pulsars; above 8 keV, polarimetry probes the origin of the hard power-law tails extending to $\sim 100$ keV.
    \item \textbf{Extragalactic jets:} AGN jets exhibit dynamic high polarization driven by structured magnetic fields near the black hole; polarimetry constrains acceleration mechanisms on resolved kpc scales.
\end{itemize}

\subsubsection*{The technical leap enabling this science}
\begin{itemize}
    \item \textbf{Advanced ASICs and 3-D track reconstruction:} Current technology already offers MEDIPIX-family ASICs (Timepix3 and the recent Timepix4) with 55-µm pitch and dead-time-free operation. Crucially, the sparse readout of Timepix3/4 provides very high time resolution of the Time of Arrival of charge at each pixel, enabling complete 3-D track reconstruction—a capability not available on IXPE. Timepix3 has already been qualified for space applications and is currently being used aboard the International Space Station (ISS) and other satellites. 
    \item \textbf{AI and sub-pixel accuracy:} Full 3-D reconstruction increases sensitivity for a more precise azimuthal-angle determination, and AI techniques in determining the impact point enable sub-pixel location accuracy. For optics with a focal length of 10 m or larger, this enables polarimetry at the 2-arcsecond level.
    \item \textbf{Broadband capabilities:} An extended band from 1.5 keV up to 30 keV requires a set of at least three devices with different gas mixtures. For energies < 1 keV, Bragg reflection polarimetry is currently being brought to TRL 6, with component testing underway to extend the bandpass up to 1 keV. The GPD lineage itself is being scaled and space-qualified by eXTP (CAS-approved, adopted mid-2025 for a ~2030 launch), which carries four polarimetry telescopes in the 2–10 keV band and inherits the PolarLight/IXPE flight heritage—thereby further reducing risk for a Lynx-class polarimeter \cite{Zhang_2025_X}.
\end{itemize}

\subsubsection{Extending Focusing to Hard X-rays}
The HEX-P Probe study advances the high-energy response, broadening Lynx's reach. A NuSTAR successor, HEX-P targets 2–200 keV with $\sim 40×\times$ the sensitivity of any previous mission in the 10–80 keV band and would be the first focusing instrument in the 80–200 keV band. Its High-Energy Telescope uses thin electroformed Ni shells coated with depth-graded Pt/C and W/Si multilayers, delivering ~10\arcsec HPD at 6 keV and ~23\arcsec at 60 keV. For Lynx, the directly transferable advance is the multilayer-coating technology: applied to a Lynx-class optic at longer focal length, such coatings enable shallower graze angles and meaningful effective area above 10 keV—extending hard-band response and, in concert with the polarimeter, reaching the reflection phenomena even in faint AGNs and binaries and cyclotron-line and hard-tail science noted above.

\subsubsection{Wider-Field and Time-Domain Detector Concepts}
Detectors that achieve even wider fields of view at full angular resolution while still resolving key X-ray transitions are of interest for wide-field targets, expanded survey science, and time-domain and multi-messenger (TDAMM) astrophysics. Microcalorimeters, while highly compelling, entail trade-offs among complexity, pixel size, field of view, and count-rate limits, so complementary detector technologies—some currently at lower TRL for spaceflight—may offer interesting trade-offs in energy resolution, to be weighed against added cost and architectural complexity.

\subsection{New Mission Architectures and Economic Strategy}

\subsubsection{New Launch and Spacecraft Capabilities}
The advent of Starship and New Glenn, offering 18–21 m fairing lengths and dramatic improvements in cost-per-mass, provides not only the potential for increased collecting area but also far-reaching architectural possibilities. These fairings allow significantly longer focal lengths without the strict need for a deployable boom, enabling shallower graze angles, improved response at harder energies, and additional space for grating dispersion. Longer focal length also relieves pressure on detector miniaturization by easing plate-scale requirements for PSF oversampling—directly benefiting the 2\arcsec polarimetry and CAT-grating cases above.

\subsubsection{Economic and Implementation Strategy}
Recognizing the Decadal mandate for cost-effective execution, the working group proposes an ambitious strategy to reduce the cost of a flagship X-ray mission toward a more manageable $\sim \$ 3$B by shifting the traditional funding profile:
\begin{itemize}
    \item \textbf{Front-loaded technology investment:} invest heavily upfront (e.g., $\sim \$$500M sustained over 5–10 years) to drive critical technologies—optics manufacturing, 100,000+ pixel microcalorimeter arrays, fine-pitch ASICs, polarimetry, and gratings—to high TRLs and Manufacturing Readiness Levels through rigorous, system-level demonstrations before formal mission development begins.
    \item \textbf{Compressed protoflight development:} by retiring innovation risk early—before the massive infrastructure of a spacecraft program begins charging to the project—the mission can be proposed as a protoflight telescope on a compressed ~5-year development timeline.
    \item \textbf{Modular servicing:} designing the optical bench and instrument paths modularly allows future robotic servicing, enabling first-generation instruments to be upgraded without building an entirely new telescope. Second-generation instruments currently at low TRL may fill capabilities not enabled by the primary suite.
\end{itemize}

\subsection{Synthesis and Outlook}
The net assessment is encouraging. The two technologies that most distinguished Lynx from its predecessors—the arcsecond, square-meter optic and the >100,000-pixel microcalorimeter—have both advanced materially since 2020. The microcalorimeter has, in effect, been flight-validated (XRISM) and scaled (LEM), and is under development towards TRL 5 with a credible manufacturing path. The CAT gratings have exceeded their Lynx requirement in the laboratory and are no longer the limiting spectroscopic element. The optic remains the pacing item: silicon meta-shell segments individually meet requirements, and AXIS/STAR-X have demonstrated module-level performance, but integrating thousands of segments to flagship scale—and the tightening required to support $R \approx 10,000$ grating spectroscopy—remains the central technical and programmatic risk.

At the same time, the science and technology landscape has handed Lynx new opportunities. X-ray polarimetry, absent from the 2020 baseline, is now a flight-proven capability (IXPE) being industrialized (eXTP) and could be transformed by a Lynx-class optic into a few-arcsecond, broadband, thousands-of-targets enterprise via 3-D-track GPDs and AI reconstruction. HEX-P demonstrates the multilayer coatings needed to push the response to hard X-rays. JWST's Little Red Dots have supplied a marquee science case—the X-ray-faint, possibly Compton-thick infant black holes of the early Universe—that a strengthened, polarimetry- and hard-band-capable Lynx is uniquely positioned to address. The case for sustained, front-loaded investment in these enabling technologies is, accordingly, stronger today than it was in 2020.

\section{Survey Grasp Working Group}
\sectionrule

In this section, we outline the conclusions of the Lynx2030 SAG Survey Grasp working group (SGWG). This work considers how the science enabled by a Lynx-like mission in the next decade could be enhanced by increasing the survey grasp (hereafter "grasp") of the telescope proposed for Lynx2020. The SGWG focused on the case for extending grasp by widening the field-of-view (FoV) of the telescope's instruments, whilst keeping the collecting area of the Lynx2020 design at its original minimum value, 2 m$^2$ effective area at 1 keV. The Lynx2020 design of the High-Definition X-ray Image (HDXI) included an array of 0.3$''$ pixels with a 22$' \times 22'$ FoV. The Lynx Mirror Assembly determined this detector size. In turn, the Lynx X-ray Microcalorimeter (LXM) was originally designed to have a Main array with a 5$' \times 5'$ FoV (1$''$ pixels) and an Enhanced Main Array in the inner part with a FoV 1$' \times 1'$ (0.5$''$ pixels). 

\subsection{Expanding the Field-Of-View of the LXM}

In \S\ref{subsec:deltas}, the Microcalorimeter Working Group discusses increasing the FoV of the LXM by adding an \emph{Outer wide-field array}, 10$'$-15$'$ on a side, with pixels $2''-5''$ in size. A $10'$ field would cover most of a galaxy's circumgalactic medium (CGM) at approximately 60 Mpc, which would enable Lynx to perform intensity mapping of some key systems to address the objectives of the \emph{Drivers of Galaxy Evolution (DGE)} Science Pillar's LXM programs. Extending the FoV of the LXM in such a way would also offer some improvements in the pursuit of the programs within the Lynx Science Pillars \emph{The Dawn of Black Holes (DBH)} and \emph{The Energetic Side of Stellar Evolution and Stellar Ecosystems (SEE)}, especially the \emph{Endpoints of stellar evolution: SNRs} program, in the latter Science Pillar. However, in most cases, the impact would be moderate given the technical necessity of lowering the angular resolution of the \emph{Outer wide-field array} to >2$''$. Studies of supernova remnants (SNRs) and other similar extended sources greatly benefit from sub-arcsecond resolution; thus, high-energy studies of objects with angular sizes larger than 5$'$ would likely rely on mosaicking.

\subsection{Drivers for extending the HDXI Field-Of-View}

Increasing the HDXI FoV, with subarcsecond angular resolution across the detector, would significantly advance several programs originally envisaged within the Lynx Science Pillars. For the survey programs in \emph{The Dawn of Black Holes (DBH)}, a larger FoV would proportionally decrease the required exposure time. While this is an "incremental" change, given that one of the Lynx performance drivers, \emph{The Origin of Supermassive Black Hole (SMBH) seeds} originally required an exposure time of 23 Msec with HDXI, even moderate expansions of the FoV would allow Lynx to complete such a program within a much shorter timeline. To illustrate this, if the HDXI FoV was extended to 30$' \times 30'$, this program would be completed in $\sim$12 Msec (a reduction of $\sim20$ weeks of continuous observation). The timeline of other surveys, such as those designed to study the X-ray binary populations in nearby galaxies, would also improve proportionally. 

The main science driver for extending the HDXI FoV is time-domain astronomy, an area that has gained significant importance since the original Lynx Case Study Report. As Rubin, Roman, and other facilities come online, along with the extraordinary results from multi-messenger detectors such as LIGO, increasing the HDXI's FoV would greatly improve its ability to contribute to time-domain studies, either by following up other observations or serendipitously detecting time-varying sources. This working group believes that a deeper study of the FoV-size requirements for Lynx to maximize its synergies with current and upcoming facilities would yield important conclusions about some of the most fundamental science results a Lynx-like mission could deliver.

\section{Credits}

\noindent
Figure on frontispiece: A conceptual X-ray flagship survey of a Coma-like cluster and its environs at $z = 0.12$, revealing Cosmic Web filaments and the surrounding large-scale structure, much of which is unobservable with current facilities. The main image shows a realistic mock of the soft X-ray surface brightness from a large cluster in the Hydrangea simulation \cite{Bahe2017}. Adapted from the Lynx Concept Study Report. The top-left inset box shows a simulated microcalorimeter image of the cluster core \cite{Ruszkowski2019}, providing spatially resolved spectroscopy to measure the thermal and kinematic properties of the region dominated by feedback from the central active galactic nucleus. The center-right inset box highlights a region in the cluster outskirts, where a microcalorimeter instrument with $\sim$eV resolution will map the radial profile of the velocity dispersion out to the virial radius \cite{Zhang2024}, revealing the kinematics of the region where matter from the Cosmic Web is accreting onto the cluster. The bottom inset highlights an otherwise bare-looking region of sky, which is filled with distant growing black holes in the early universe and nearby galaxy groups. For more nearby active galactic nuclei within the field, microcalorimeter and grating instruments will provide a detailed view of the accretion and feedback processes in these systems, enabling a comprehensive understanding of the interplay between galaxies and their environments (simulated spectra based on the model of the quasar PDS-456 based on XRISM observations by \cite{XRISM_PDS456_2025,Xu2025}). Also within this region, a microcalorimeter-based velocity map of a simulated M82-like starburst galaxy from the Cholla Galactic Outflow Simulation Suite \cite{Schneider2015,Schneider2018a,Schneider2018b} reveals the rotation and outflow velocities of the hot gas.

\clearpage
\phantomsection
\addcontentsline{toc}{section}{References}
\bibliographystyle{aasjournal}
\bibliography{references}

\end{document}